\documentclass[11pt]{article}

\newcommand{\ifabs}[2]{#2}

\usepackage{epsf}
\usepackage{citesort}
\usepackage{epsfig}

\newtheorem{theorem}{Theorem}
\newtheorem{lemma}[theorem]{Lemma}

\newenvironment{proof}{\noindent\par{\bf Proof: }}{\nopagebreak\rule{1
ex}{0.8 em}\medskip}

\newcommand{\SSS}{{\cal X}}
\newcommand{\f}{\Phi}
\newcommand{\hatf}{\Phi'}
\newcommand{\FF}{\overline{\Phi}}
\newcommand{\FFT}{F}
\newcommand{\DD}{{\cal D}}
\newcommand{\GG}{{\cal A}}
\newcommand{\KK}{{\cal B}}
\newcommand{\LL}{{\rm line}}
\newcommand{\angstroms}{\AA}
\newcommand{\sharpP}{\mbox{\#P}}
\newcommand{\NP}{\mbox{NP}}
\newcommand{\Z}{\mbox{\bf Z}}
\newcommand{\R}{\mbox{\bf R}}

\newcommand{\buzz}[1]{\emph{#1}}
\newcommand{\etal}{\emph{et.~al.}}
\newcommand{\opt}[1]{{\rm opt}(#1)}
\newcommand{\mopt}[1]{{\mathbf\mathrm opt}(#1)}

\newcommand{\fsetupm}{Let $\f(x) = -\sum_{1\leq i < j \leq n} a_{i,j}
x_i x_j + \sum_{1 \leq i \leq n} b_i x_i$, where $a_{i,j} \geq 0$,
$b_i$ is arbitrary, and $m$ of the coefficients $a_{i,j}$ are
nonzero. Let $\Delta = n+m$.}

\newcommand{\fsetuptune}{For each $\beta \geq 0$, let $\f_{\beta}(x) =
- \sum_{1 \leq i < j \leq n} a_{i,j} x_i x_j + \beta \sum_{1 \leq i
\leq n} s_i x_i,$ where $a_{i,j} \geq 0$, $s_i \geq 0$, and $m$ of the
coefficients $a_{i,j}$ are nonzero. Let $\Delta = n+m$.}

\newcommand{\fsetupmulti}{For each $\ell$ from $1$ to $f$, let the
$\ell$-th fitness function $\f^\ell(x) = -\sum_{1\leq i < j \leq n}
a^\ell_{i,j} x_i x_j + \sum_{1 \leq i \leq n} b^\ell_i x_i$, where 
$a^\ell_{i,j} \geq 0$ and $b^\ell_i$ is arbitrary. Let $\Delta=fn^2$.}

\newcommand{\ftwo}{Let $\f$ be as defined in Assumption~\ref{f2}.}
\newcommand{\fthr}{Let $\f^1,\ldots,\f^f$ be as defined in Assumption~\ref{f3}.}
\newcommand{\fthrr}{$\f^1,\ldots,\f^f$ are as defined in Assumption~\ref{f3}}
\newcommand{\ffou}{Let $\f$ be as defined in Assumption~\ref{f4}.}

\newcommand{\minCutTime}{\Delta^2 \log \Delta}

\newcommand{\Gfsta}{G^{\f}_{s,t}}
\newcommand{\Gfst}{\widehat{G}^{\f}_{s,t}}
\newcommand{\Gfstxa}[1]{G^{\f^{#1}}_{s,t}}
\newcommand{\Gfstx}[1]{\widehat{G}^{\f^{#1}}_{s,t}}
\newcommand{\Gfstla}{\Gfstxa{\ell}}
\newcommand{\Gfstl}{\Gfstx{\ell}}
\newcommand{\Gfmultia}{\Gfstxa{*}}
\newcommand{\Gfmulti}{\Gfstx{*}}

\newcommand{\givenGfst}{Given the $\Gfsta$ and $\rho$ 
defined in Theorem~\ref{theorem-f-to-graph}}

\newcommand{\isom}{\cong}
\newcommand{\symdiff}{\ominus}

\newcommand{\eqLP}{Linear Program $(\ref{eq-LP})$}

\newcommand{\dobib}{\bibliographystyle{abbrv}
\bibliography{all}}

\newcommand{\acknowledgments}{
\section*{Acknowledgments}
We wish to thank Jon Kleinberg for generously providing \LaTeX{}
entries for many of the references; thank Lisa Fleischer and Hal Gabow
for their help in tracking down references used to generalize
Theorem~\ref{theorem-f-to-graph} for submodular functions; and thank
Mark Gerstein for helpful discussions.}

\begin{document}

\title{A Combinatorial Toolbox for Protein Sequence Design and
Landscape Analysis in the Grand Canonical Model}

\author{
James Aspnes\thanks{Department of Computer Science, Yale University,
New Haven, CT 06520-8285, USA.
Email: {\tt aspnes@cs.yale.edu}.
Supported in part by NSF Grant CCR-9820888.} 
\and
Julia Hartling\thanks{Department of Ecology and Evolutionary Biology,
Yale University, New Haven, CT 06520-8285, USA.  Email: {\tt
julia.kreychman@yale.edu}.}
\and
Ming-Yang Kao\thanks{Department of Electrical Engineering and Computer Science,
Tufts University, Medford, MA 02155, USA. Email: {\tt kao@eecs.tufts.edu}.
Supported in part by NSF Grant CCR-9531028.}
\and
Junhyong Kim\thanks{
Department of Ecology and Evolutionary Biology,
Department of Molecular, Cellular, and Developmental Biology, and 
Department of Statistics, 
Yale University, New Haven, CT 06520-8285, USA.  Email: {\tt
junhyong.kim@yale.edu}.
Supported in part by 
Merck Genome Research Institute Grant and NSF Grant DEB-9806570.
}
\and
Gauri Shah\thanks{Department of Computer Science, Yale University, New
Haven, CT 06520-8285, USA.  Email: {\tt gauri.shah@yale.edu}.}}

\maketitle

\begin{abstract}
In modern biology, one of the most important research problems is to
understand how protein sequences fold into their native 3D structures.
To investigate this problem at a high level, one wishes to analyze the
{\it protein landscapes}, i.e., the structures of the space of all
protein sequences and their native 3D structures. Perhaps the most
basic computational problem at this level is to take a target 3D
structure as input and {\it design} a fittest protein sequence with respect to
one or more fitness functions of the target 3D structure.  We develop a
toolbox of combinatorial techniques for protein landscape analysis in
the Grand Canonical model of Sun, Brem, Chan, and Dill.  The toolbox
is based on linear programming, network flow, and a linear-size
representation of all minimum cuts of a network. It not only
substantially expands the network flow technique for protein sequence
design in Kleinberg's seminal work but also is applicable to a
considerably broader collection of computational problems than those
considered by Kleinberg. We have used this toolbox to obtain a number
of efficient algorithms and hardness results.  We have further used
the algorithms to analyze 3D structures drawn from the Protein Data Bank and
have discovered some novel relationships between such native 3D structures and
the Grand Canonical model.
\end{abstract}

\newpage

\section{Introduction} 
In modern biology, one of the most important research problems is to
understand how protein sequences fold into their native 3D structures
\cite{Merz:1994:PFP}.  This problem can be investigated at two complementary
levels.  At a low level, one wishes to determine how an individual
protein sequence folds.  A fundamental computational problem at this
level is to take a protein sequence as input and find its
native 3D structure.  This problem is sometimes referred to as the
protein {\it structure prediction} problem and has been shown to be
NP-hard (see, e.g., \cite{Atkins:1999:IPF,Crescenzi:1998:CPF,Berger:1998:PFH}).  At a high level, one wishes
to analyze the {\it protein landscapes}, i.e., the structures of the
space of all protein sequences and their native 3D structures. Perhaps
the most basic computational problem at this level is to take a target
3D structure as input and ask for a fittest protein sequence with respect to
one or more fitness functions of the target 3D structure.  This problem
has been called the protein {\it sequence design} problem and has been
investigated in a number of studies
\cite{Drexler:1981:MEA,Ponder:1987:TTP,Yue:1992:IPF,Shakhnovich:1993:NAD,Deutsch:1996:NAP,Sun:1995:DAA,Hart:1997:CCS,Banavar:1998:SBD}.

The focus of this paper is on protein landscape analysis, for which
several quantitative models have been proposed in the literature
\cite{Sun:1995:DAA,Shakhnovich:1993:NAD,Deutsch:1996:NAP}.  As some recent studies on this topic
have done \cite{Micheletti:1998:DPH,Kleinberg:1999:EAP,Banavar:1998:SBD}, this paper employs
the Grand Canonical (GC) model of Sun, Brem, Chan, and
Dill~\cite{Sun:1995:DAA}, whose definition is given in
Section~\ref{sec_gc}. Generally speaking, the model is specified by
(1) a 3D geometric representation of a target protein 3D structure with
$n$ amino acid residues, (2) a {\it binary folding code} in which the
amino acids are classified as {\it hydrophobic} (H) or {\it polar} (P)
\cite{Dill:1995:PPF,Lau:1989:LSM}, and (3) a fitness function $\f$ defined
in terms of the target 3D structure that favors protein sequences with a dense
hydrophobic core and with few solvent-exposed
hydrophobic residues.

In this paper, we develop a toolbox of combinatorial techniques for
protein landscape analysis based on linear programming, network flow,
and a linear-size representation of all minimum cuts of a network
\cite{Picard:1980:SAM}.  This toolbox not only substantially expands the
network flow technique for protein sequence design in Kleinberg's
seminal paper \cite{Kleinberg:1999:EAP} but also is applicable to a
considerably broader collection of computational problems than those
considered by Kleinberg. We have used this toolbox to obtain a number
of efficient algorithms and hardness results.  We have further used
the algorithms to analyze 3D structures drawn from Protein Data Bank
at \texttt{http://www.rcsb.org/pdb} and have discovered some novel
relationships between such native 3D structures and the Grand Canonical
model (Figure~\ref{fig:similarity_PFAM}).  Specifically, we report new
results on the following problems, where $\Delta$ is the number of
terms in the fitness function or functions as further defined in
Section~\ref{sec_linear_prog}.  Many of the results depend on
computing a maximum network flow in a graph of size $O(\Delta)$; in
most cases, this network flow only needs to be computed once for each
fitness function $\f$.

\newcounter{hgt}
\begin{list}{\thehgt}
{\usecounter{hgt}\setcounter{hgt}{0}\renewcommand{\thehgt}{P\arabic{hgt}}
\setlength{\rightmargin}{0in} 
\settowidth{\leftmargin}{P12} \addtolength{\leftmargin}{\labelsep}}
\item\label{prob_all_sequences} Given a 3D structure, find all its
fittest protein sequences.  Note that there can be exponentially many fittest
protein sequences. We show that these protein sequences together have a representation
of size $O(\Delta)$ that can be computed in
$O(\Delta)$ time after a certain maximum network flow is computed
(Theorem~\ref{theorem-f-to-graph}), and
that individual fittest protein sequences can be generated from this
representation in $O(n)$ time per sequence 
(Theorem~\ref{theorem-enumerating-optimal}).
\item\label{prob_p2} Given $f$ 3D structures, find the set of all
protein sequences that are the fittest simultaneously for all these
3D structures.  This problem takes $O(\Delta)$ time after $f$ 
maximum network flow computations
(Theorem~\ref{theorem-intersection}).
\item \label{prob_p3} Given a protein sequence $\hat{x}$ and its native 3D
structure, find the set of all fittest protein sequences that are also the
most (or least) similar to $\hat{x}$ in terms of unweighted (or weighted)
Hamming distances.  This problem takes $O(\Delta)$ time after a
certain maximum network flow is computed
(Theorem~\ref{thm_dist_Hcount}).
\item \label{prob_p5} Count the number of protein sequences in the solution to
each of Problems~\ref{prob_all_sequences}, \ref{prob_p2}, and
\ref{prob_p3}.  These counting problems are computationally hard
(Theorem~\ref{theorem-counting-optimal-is-hard}).
\item \label{prob_p6} Given a 3D structure and a bound $e$, enumerate
the protein sequences whose fitness function values are within an additive
factor $e$ of that of the fittest protein sequences. This problem takes
polynomial time to generate each desired protein sequence
(Theorem~\ref{theorem-enumerating-suboptimal}).
\item \label{prob_p4} Given a 3D structure, determine the largest
possible unweighted (or weighted) Hamming distance between any two
fittest protein sequences. This problem takes $O(\Delta)$ time after a certain
maximum network flow is computed
(Theorem~\ref{theorem-diameter}).
\item \label{prob_p8} Given a protein sequence $\hat{x}$ and its native 3D
structure, find the average unweighted (or weighted) Hamming distance
between $\hat{x}$ and the fittest protein sequences for the 3D structure.  This
problem is computationally hard
(Theorem~\ref{theorem-counting-optimal-is-hard}).
\item \label{prob_p10} Given a protein sequence $\hat{x}$, its native
3D structure, and two unweighted Hamming distances $d_1$ and $d_2$,
find a fittest protein sequence whose distance from $\hat{x}$ is also
between $d_1$ and $d_2$.  This problem is computationally hard
(Theorem~\ref{theorem-bounded-distance-is-hard}(\ref{thm_bound_d_1})).
\item \label{prob_p9} Given a protein sequence $\hat{x}$, its native
3D structure, and an unweighted Hamming distance $d$, find the fittest
among the protein sequences which are at distance $d$ from $\hat{x}$.
This problem is computationally hard
(Theorem~\ref{theorem-bounded-distance-is-hard}(\ref{lem_dist_d})). We
have a polynomial-time approximation algorithm for this problem
(Theorem~\ref{theorem-enumerating-by-Hamming-distance}).
\item \label{prob_p7} Given a protein sequence $\hat{x}$ and its
native 3D structure, find all the ratios between the scaling factors
$\alpha$ and $\beta$ in Equation \ref{eq-phi-definition} in Section
\ref{sec_gc} for the GC model such that the smallest possible
unweighted (or weighted) Hamming distance between $\hat{x}$ and any
fittest protein sequence is minimized over all possible $\alpha$ and
$\beta$. (This is a problem of tuning the GC model.)  We have a
polynomial-time algorithm for this problem
(Theorem~\ref{theorem-tuning}).
\item\label{prob_all_connected} Given a 3D structure, determine
whether the fittest protein sequences are {\it connected}, i.e., whether they
can mutate into each other through allowable mutations, such as point
mutations, while the intermediate protein sequences all remain the fittest
\cite{Smith:1970:NSC,Dill:1995:PPF,Lau:1990:TPM,Lipman:1991:MNS,Babajide:1997:NNP,Kimura:1983:NTM,Reidys:1997:GPC}.
This problem takes $O(\Delta)$ time after a certain maximum network
flow is computed
(Theorem~\ref{theorem-connectivity}).
\item\label{prob_two_connected} Given a 3D structure, in the case that
the set of all
fittest protein sequences is not connected, determine whether
two given fittest protein sequences are connected.  This problem takes
$O(\Delta)$ time after a certain maximum network flow is computed
(Theorem~\ref{theorem-connectivity}).
\item\label{prob_connected_mutation} Given a 3D structure, find the
smallest set of allowable mutations with respect to which the fittest
protein sequences (or two given fittest protein sequences) are connected.  This
problem takes $O(\Delta)$ time after a certain maximum network flow is
computed
(Theorem~\ref{theorem-connectivity}).
\end{list}

Previously, Sun~\etal~\cite{Sun:1995:DAA} developed a heuristic
algorithm to search the space of protein sequences for a fittest protein sequence
without a guarantee of optimality or near-optimality.  Hart
\cite{Hart:1997:CCS} subsequently raised the computational tractability of
constructing a single fittest protein sequence as an open question. Kleinberg
\cite{Kleinberg:1999:EAP} gave the first polynomial-time algorithm for this
problem, which is based on network flow.  In contrast,
Problem~\ref{prob_all_sequences} asks for all fittest protein sequences and
yet can be solved with the same time complexity.  Kleinberg also
formulated more general versions of Problems~\ref{prob_all_connected}
and \ref{prob_two_connected} by extending the fitness function to a
submodular function and gave polynomial-time algorithms.  Our
formulations of these two problems and
Problem~\ref{prob_connected_mutation} are directly based on the
fitness function of the GC model; furthermore, as is true with several
other problems above, once a solution to
Problem~\ref{prob_all_sequences} is obtained, we can solve these three
problems in $O(\Delta)$ time. Among the above thirteen problems, those
not yet mentioned in this comparison were not considered by Kleinberg.

The remainder of this paper is organized as follows.
Section~\ref{sec_gc} defines the GC model and states the basic
computational assumptions.
Section~\ref{section-basic-tools} describes our three basic tools
based on linear programming, network flow, and an $O(\Delta)$-size
representation of minimum cuts.  
Section~\ref{section-further-tools} extends these tools to optimize
multiple objectives, analyze the structures of the space of all
fittest protein sequences, and generate near-fittest
protein sequences. Section~\ref{section-hard-problems} gives some hardness
results related to counting fittest protein sequences 
and finding fittest protein sequences under additional restrictions.
\ifabs{In the full paper, we apply our techniques to analyze empirical
3D structures from the Protein Data Bank.  For reasons of space, in this
extended abstract 
discussion of this analysis is deferred to the appendix.
}{Finally, Section~\ref{section-pdb} discusses
our analysis of empirical 3D structures from the Protein Data Bank.
}

\ifabs{We also omit proofs of our results;
these proofs appear in the full paper.}

\section{The Grand Canonical Model and Computational Assumptions}
\label{sec_gc}

\paragraph{The Original Model}
Throughout this paper, all protein sequences are of $n$ residues,
unless explicitly stated otherwise.  The GC model is specified by a
fitness function $\f$ over all possible protein sequences $x$ with
respect to a given 3D structure of $n$ residues
\cite{Kleinberg:1999:EAP,Sun:1995:DAA}.  In the model, to design a protein
sequence $x$ is to specify which residues are hydrophobic ($H$) and
which ones are polar ($P$).  Thus, we model $x$ as a binary sequence
$x_1,\ldots,x_n$ or equivalently as a binary vector
$(x_1,\ldots,x_n)$, where the $i$-th residue in $x$ is $H$
(respectively, $P$) if and only if $x_i = 1$ (respectively, $0$).
Then, $\f(x)$ is defined as follows, where the smaller $\f(x)$ is, the
fitter $x$ is, as the definition is motivated by the requirements that
$H$ residues in $x$ (1) should have low solvent-accessible surface
area and (2) should be close to one another in space to form a compact
hydrophobic core.

\begin{eqnarray}
\label{eq-phi-definition}
\f(x) &=& \alpha\sum_{i,j \in H(x), i < j-2} g(d_{i,j}) 
          + \beta\sum_{i \in H(x)} s_i 
\\  
\label{eq-phi-vs-x}
      &=& \alpha\sum_{i < j-2} g(d_{i,j})x_i x_j 
          + \beta\sum_i s_i x_i,\ \mbox{where} 
\end{eqnarray}
\begin{itemize}
\item 
$H(x) = \{ i \mid x_i=1 \}$,
\item 
the scaling parameters $\alpha < 0$ and $\beta > 0$ have default
values $-2$ and $\frac{1}{3}$ respectively and may require tuning for
specific applications (see Section~\ref{section-tuning}),
\item 
$s_i \geq 0$ is the area of the solvent-accessible contact surface for
the residue (in \angstroms) \cite{Eisenhaber:1995:DCL,Eisenhaber:1993:ISA},
\item 
$d_{i,j} > 0$ is the distance between the residues $i$ and $j$ (in
\angstroms), and
\item $g$ is a sigmoidal function, defined by
\begin{displaymath}
g =
\left\{
\begin{array}{cl}
\frac{1}{1 + \exp(d_{i,j} - 6.5)} & \mbox{when $d_{i,j} \le 6.5$} \\
0 & \mbox{when $d_{i,j} > 6.5$.}
\end{array}
\right.
\end{displaymath}
\end{itemize}

\paragraph{Extending the Model with Computational Assumptions} 
Let $\opt{\f}$ be the set of all protein sequences $x$ that minimize $\f$.
This paper is generally concerned with the structure of $\opt{\f}$.
Our computational problems assume that $\f$ is given as input; in
other words, the computations of $\alpha, \beta, s_i, g(d_{i,j})$ are
not included in the problems.  Also, for the sake of computational
generality and notational simplicity, we assume that $\alpha$ may be
any nonpositive number, $\beta$ any nonnegative number, $s_i$ any
arbitrary number, and $g(d_{i,j})$ any arbitrary nonnegative number;
and that the terms $g(d_{i,j})$ may range over $1 \leq i < j \leq n$, unless
explicitly stated otherwise. Thus, in the full generality of these
assumptions, $\f$ need not correspond to an actual protein 3D
structure.  Note that the relaxation that $s_i$ is any number is
technically useful for finding $\f$-minimizing protein sequences $x$ that satisfy
additional constraints.

We write $a_{i,j} = -\alpha{\cdot}g(d_{i,j}) \geq 0$ and $b_i =
\beta{\cdot}s_i$ and further assume that the coefficients $a_{i,j}$ and
$b_i$ are rational with some common denominator, that these
coefficients are expressed with a polynomial number of bits, and that
arithmetic operations on these coefficients take constant time.

With these assumptions, we define the following sets of specific
assumptions about $\f$ to be used at different places of this paper.

\newcounter{fitness}
\begin{list}{\thefitness}
{\usecounter{fitness}\setcounter{fitness}{0}\renewcommand{\thefitness}{F\arabic{fitness}}
\setlength{\rightmargin}{0in} 
\settowidth{\leftmargin}{F3} \addtolength{\leftmargin}{\labelsep}}
\item\label{f2} \fsetupm{}
\item\label{f4} \fsetuptune{}
\item\label{f3} \fsetupmulti{}
\end{list}

Sometimes we measure the dissimilarity between a fittest protein
sequence $x$ and a target protein sequence $\hat{x}$ in terms of
Hamming distance. This distance is essentially the count of the
positions $i$ where $x_i\not=\hat{x}_i$ and can be measured in two
ways.  The {\it unweighted} Hamming distance is $|x-\hat{x}|$, where
$|y|$ denotes the {\it norm} of vector $y$, i.e., $\sum_{i=1}^n
|y_i|$.  The {\it weighted} Hamming distance is $\sum_{i=1}^n
w_i{\cdot}|x_i-\hat{x}_i|$. Throughout this paper, the weights
$w_1,\ldots,w_n$ are all arbitrary unless explicitly stated otherwise.

\section{Three Basic Tools}
\label{section-basic-tools}
This section describes our basic tools for computing fittest and
near-fittest protein sequences. For instance, Lemma~\ref{lemma-LP}
gives a representation of the problem of minimizing $\f$ as a linear
program.  Lemma~\ref{lemma-cuts} further gives a representation of
this problem as a minimum-cut problem, which generalizes a similar
representation of Kleinberg~\cite{Kleinberg:1999:EAP}.
Theorem~\ref{theorem-f-to-graph} gives a compact representation of the
space $\opt{\f}$ using a Picard-Queyranne graph~\cite{Picard:1980:SAM}.

\subsection{Linear Programming}\label{sec_linear_prog}
{From} Equation~\ref{eq-phi-vs-x}, minimizing $\f(x)$ is an optimization
problem in quadratic programming.  Fortunately, because all the
coefficients $a_{i,j}$ are nonnegative, it can be converted to a linear
program, as shown in Lemma~\ref{lemma-LP}.

\begin{lemma}[characterizing $\mathbf{\f}$ via linear program]
\label{lemma-LP}
\ftwo{}
Consider the following linear program whose variables consist of the
variables $x_i$, together with new variables $y_{i,j}$
for all $i,j$ with $a_{i,j} \neq 0$:
\begin{equation}
\label{eq-LP}
\begin{array}{l}
\mbox{minimize}\  
\hatf(x,y) = 
-\sum a_{i,j} y_{i,j} + \sum b_i x_i\\
\mbox{subject to} \\
\begin{array}{cl}
\!\!\!\!\! 0 \le x_i \le 1 & \forall i \\
\left.
\begin{array}{c}
0 \le y_{i,j} \le 1 \\
y_{i,j} \le x_i 	\\
y_{i,j} \le x_j \\
\end{array}
\right\}
&
\forall i,j: a_{i,j} \neq 0\\
\end{array}
\end{array}
\end{equation}
There is a one-to-one correspondence that preserves $x$ between the
protein sequences that minimize $\f(x)$ and the basic optimal
solutions to \eqLP{}.
\end{lemma}
\ifabs{}{\begin{proof}
First, we show that for each $0$-$1$ assignment to $x$ there is
a unique value of $y$ that minimizes $\hatf(x,y)$.  
Choose some $y_{i,j}$, and suppose that either $x_i$ or $x_j$
is $0$.  Then $y_{i,j}$ is also $0$ by the 
constraint $y_{i,j} \le x_i$ or $y_{i,j} \le x_j$.
Alternatively, suppose $x_i$ and $x_j$ are both $1$; then if $y_{i,j}$
is $0$, $\hatf$ can be decreased by $a_{i,j}$ by setting
$y_{i,j}$ to $1$ without violating any constraints. 
Thus, in any optimal integral solution to \eqLP{},
$y_{i,j} = \min(x_i,x_j) = x_i x_j$.

Note that substituting $x_i x_j$ for  $y_{i,j}$ in $\hatf$
gives precisely 
$-\sum_{i,j} a_{i,j} x_i x_j + \sum_i b_i x_i = \f(x)$; thus minimizing
$\hatf(x,y)$
is equivalent to minimizing $\f(x)$.

We now must show that all solutions to \eqLP{} are integral.  Every
element of the constraint matrix is either zero or $\pm 1$.  Each row
has either a single nonzero element (e.g, for the $0$-$1$ bounds) or
consists of zeroes and exactly one $+1$ and one $-1$.  Thus the matrix
is totally unimodular, e.g., using \cite[Theorem
13.3]{Papadimitriou:1982:COA}.  Since the right-hand side is integral, any
vertex of the polytope defined by \eqLP{} is integral~\cite[Theorem
13.2]{Papadimitriou:1982:COA}.  Thus, all basic feasible solutions to
\eqLP{} are $0$-$1$ vectors.

So if $(x,y)$ is a basic optimal solution to \eqLP{}, then
$x\in\opt{\f}$.  Conversely, if $x\in\opt{\f}$, then the vector
$(x,y)$ in which $y_{i,j} = x_i x_j$ whenever $a_{i,j}$ is nonzero is an
optimal solution to \eqLP{}, which is a basic optimal solution since
an appropriate subset of the constraints $0 \le x_i \le 1$ and $0 \le
y_{i,j} \le 1$ form a basis.
\end{proof}}

Note that any $x_i$ with a
negative coefficient $b_i$ is set to $1$ in any optimal solution,
as in this case all terms containing $x_i$ have negative coefficients
and are minimized when $x_i = 1$.  So an alternative to allowing
negative coefficients is to prune out any $x_i$ with a negative
coefficient.  This process must be repeated recursively, since setting
$x_i$ to $1$ reduces terms of the form $-a_{i,j} x_i x_j$ to $-a_{i,j}
x_j$, and may yield more degree-1 terms with negative coefficients.  
To simplify our discussion, 
we let the linear program (or,
in Section~\ref{section-network-flow}, the minimum-cut algorithm) handle this
pruning.

\subsection{Network Flow}
\label{section-network-flow}
Recall that an $s$-$t$ {\it cut} is a partition of the nodes of a
digraph into two sets $V_s$ and $V_t$, with $s\in V_s$ and $t\in
V_t$. Also, a {\it minimum} $s$-$t$ cut is an $s$-$t$ cut with the
smallest possible total capacity of all edges from nodes in $V_s$ to
nodes in $V_t$.  

In Kleinberg's original construction~\cite{Kleinberg:1999:EAP}, $\f(x)$ was
minimized by solving an $s$-$t$ minimum cut problem in an appropriate
digraph $G$.  Lemma~\ref{lemma-cuts} describes a more general
construction that includes additional edges $(s,v_i)$ to handle
negative values for $b_i$.

\begin{lemma}[characterizing $\mathbf{\f}$ via network flow]
\label{lemma-cuts}
\ftwo{} Let $G^{\f}$ be a graph with a source node $s$, a sink node
$t$, a node $v_i$ for each $i$, and a node $u_{i,j}$ for each $i,j$
with $a_{i,j} \not= 0$, for a total of $n+m+2 = \Delta+2$ nodes.  Let
the edge set of $G^{\f}$ consist of
\begin{itemize}
\item $(s, u_{i,j})$ for each $u_{i,j}$, with capacity $a_{i,j}$,
\item $(v_i, t)$ for each $v_i$ with $b_i > 0$, 
with capacity $b_i$, 
\item $(s, v_i)$ for each $v_i$ with $b_i < 0$,
with capacity $-b_i$, and
\item $(u_{i,j}, v_i)$ and $(u_{i,j}, v_j)$, for each $u_{i,j}$,
with infinite capacity,
\end{itemize}
for a total of $\Theta(\Delta)$ edges.

There is a one-to-one correspondence between the minimum $s$-$t$ cuts
in $G^{\f}$ and the protein sequences in $\opt{\f}$, such that $v_i$
is in the $s$-component of a cut if and only if $x_i = 1$ in the
corresponding protein sequence.
\end{lemma}
\ifabs{}{\begin{proof} We will show that the minimum $s$-$t$ cuts in
$G^{\f}$ correspond to $\f$-minimizing protein sequences via \eqLP{}
of Lemma~\ref{lemma-LP}.  Given a minimum $s$-$t$ cut in $G$, let
$x_i$ be 1 if $v_i$ is in the $s$ component, and 0 otherwise.
Similarly, let $y_{i,j}$ be 1 if $u_{i,j}$ is in the $s$ component,
and $0$ otherwise.  Since no infinite-capacity edge $(u_{i,j}, v_i)$
or $(u_{i,j}, v_j)$ can appear in the cut, if $u_{i,j}$ is in the
$s$-component then $v_i$ and $v_j$ are as well.  In terms of the $x$
and $y$ variables, we have $y_{i,j} \le x_i$ and $y_{i,j} \le x_j$
whenever $a_{i,j}$ is nonzero, precisely the same constraints as in
\eqLP{}.  Conversely, any $0$-$1$ assignment $(x,y)$ for which these
constraints hold defines an $s$-$t$ cut that does not include any
infinite-capacity edge.

Turning to the objective function, the total capacity of all edges in
the cut is
\begin{eqnarray*}
\lefteqn{\sum_{a_{i,j} \neq 0} a_{i,j} (1-y_{i,j})
+ \sum_{b_i > 0} b_i x_i
+ \sum_{b_i < 0} -b_i (1 - x_i)} \\
& &  =  
\sum_{i,j} a_{i,j}
- \sum_{b_i < 0} b_i 
- \sum_{i,j} a_{i,j} y_{i,j}
+ \sum_{i} b_i x_i 
= K + \hatf(x,y),
\end{eqnarray*}
where $K$ is a constant and $\hatf(x,y)$ 
is the objective function of \eqLP{}.
Thus, the capacity of the cut is
minimized when $\hatf(x,y)$ is.  The rest follows from
Lemma~\ref{lemma-LP}.
\end{proof}}

\begin{lemma}
\label{lemma-minimize-cost}
\ftwo{} Given $\f$ as the input, we can find an $x \in \opt{\f}$ in
$O(\minCutTime)$ time.
\end{lemma}
\ifabs{}{
\begin{proof} 
Given a digraph $G=(V,E)$ as input, the Goldberg-Tarjan
maximum-flow algorithm takes $O(|V||E|\log(|V|^2/|E|))$ time
\cite{Goldberg:1988:NAM}.  We first apply Lemma~\ref{lemma-cuts} to $\f$ to
obtain $G^{\f}$. We next use this maximum-flow algorithm to find a
minimum $s$-$t$ cut in $G^{\f}$ and then an optimal $x$ from this cut.
All these steps take $O(\minCutTime)$ total time.
\end{proof}}

\subsection{A Compact Representation of Minimum Cuts}

A given $\f$ may have more than one fittest protein sequence.
Theorem~\ref{theorem-f-to-graph} shows that $\opt{\f}$ can be
summarized compactly using the Picard-Queyranne representation of the
set of all minimum $s$-$t$ cuts in a digraph $G$ \cite{Picard:1980:SAM},
which is computed by the following steps:
\begin{enumerate}
\item 
computing any maximum flow $\phi$ in $G$;
\item 
computing strongly connected components in the residual
graph $G_\phi$ whose edge set consists of all edges in $G$ that are not
saturated by $\phi$, plus edges $(v,u)$ for any edge $(u,v)$ that has
nonzero flow in $\phi$; 
\item  
contracting $G_\phi$ by contracting
into single supernodes
the set of all nodes reachable from $s$, 
the set of all nodes that can reach $t$,
and each strongly connected component in the remaining graph.
\end{enumerate}

The resulting graph $G_{s,t}$ is a dag in which $s$ and $t$
are mapped to distinct supernodes by the contraction.
Furthermore, there is a one-to-one correspondence between the minimum 
$s$-$t$ cuts
in $G$ and the ideals in $G_{s,t}$, where an {\it ideal} is any node set $I$
with the property that any predecessor of a node in $I$ is also in $I$.

\begin{lemma}[see \cite{Picard:1980:SAM}]
\label{lemma-PQ}
Given a digraph $G$ with designated nodes $s$ and $t$, there is
a graph $G_{s,t}$ together with a mapping $\kappa$ from $V(G)$
to $V(G_{s,t})$ with the following properties:
\begin{enumerate}
\item $|V(G_{s,t})| \le |V(G)|$.
\item The node $\kappa(s)$ has out-degree $0$ while $\kappa(t)$ has in-degree
$0$.
\item Given $G$ as the input, 
$G_{s,t}$ and $\kappa$ can be computed using one maximum-flow computation and
$O(|E(G)|)$ additional work. 
\item A partition $(V_s,V_t)$ of $V(G)$
is an $s$-$t$ minimum cut in $G$ if and only if 
$V_t = \kappa^{-1}(I)$ for some ideal $I$ of $G_{s,t}$ that contains
$\kappa(t)$ but not $\kappa(s)$.
\end{enumerate}
\end{lemma}

Combining Lemmas~\ref{lemma-cuts} and~\ref{lemma-PQ} gives the desired compact
representation of the space of all fittest protein sequences, 
as stated in the next theorem.

\begin{theorem}[characterizing $\mathbf{\f}$ via a dag]
\label{theorem-f-to-graph}
\ftwo{} There exists a dag $\Gfsta$ with designated nodes $s'$ and
$t'$ and a mapping $\rho$ from $\{1,\ldots,n\}$ to $V(\Gfsta)$ with
the following properties:
\begin{enumerate}
\item $\Gfsta$ has at most $n+2$ nodes.
\item 
Given $\f$ as the input, 
$\Gfsta$ and $\rho$ can be computed in $O(\minCutTime)$ time.
\item There is a one-to-one correspondence between the protein
sequences $x\in\opt{\f}$ and the ideals of $\Gfst = \Gfsta -
\{s',t'\}$, in which $x_i = 0$ if and only if $\rho(i) = t'$ or $\rho(i)$
is in the ideal corresponding to $x$.
\end{enumerate}
\end{theorem}
\ifabs{}{\begin{proof}
The graph $\Gfsta$ is obtained by applying
Lemmas~\ref{lemma-cuts} and~\ref{lemma-PQ}.
Let $\kappa$ be the contraction map from Lemma~\ref{lemma-PQ}.
Let $s'=\kappa(s)$ and $t'=\kappa(t)$.
The mapping 
$\rho(i)$ is defined as $\kappa(v_i)$.

To show that $\Gfsta$ has at most $n+2$ nodes, 
consider any node $u_{i,j}$ in $G^{\f}$.  Let $\phi$ be the maximum flow
used to define $\Gfsta$.  If $\phi(s,u_{i,j}) = 0$, then $u_{i,j}$ is
reachable from $s$ in the residual graph $G^{\f}_\phi$, and $u_{i,j}$ is
contracted onto the $\kappa(s)$ supernode; if
$\phi(s,u_{i,j}) > 0$, then at least one of $\phi(u_{i,j},v_i)$ or
$\phi(u_{i,j},v_j)$ is nonzero and $u_{i,j}$ is in the same
strongly connected component in $G^{\f}_\phi$ as at least one of $v_i$
and $v_j$.  In either case $u_{i,j}$ is contracted onto a supernode
that contains $s$ or some $v_i$; since the same thing happens to all
$u_{i,j}$, there are at most $n+2$ supernodes in $\Gfsta$: 
one for each $v_i$, plus
one for each of $s$ and $t$.

Using ideals of $\Gfst$ is justified by the observation that requiring
$t'$ to be in an ideal and $s'$ to be out of it has no effect on the
presence or absence of other nodes, as $t'$ has no predecessors and $s'$
has no successors in $\Gfsta$; thus there is a one-to-one
correspondence preserving all nodes except $s'$ and $t'$ between the
ideals of $\Gfsta$ containing $t'$ but not $s'$ and the ideals of
$\Gfst$.
\end{proof}}

{\it Remark.}  At some additional cost in time, Assumption~\ref{f2}
can be replaced in Theorem~\ref{theorem-f-to-graph} by the weaker
assumption that $\f$ is submodular (i.e., that $\f(X \cup Y) + \f(X
\cap Y) \le \f(X) + \f(Y)$ for all $X,Y$, where each protein sequence
$x$ in $\f$'s domain is regarded as the set $X = \{i \mid x_i = 1\}$
for the purposes of taking unions and intersections).  The reason is
that a representation similar to the Picard-Queyranne graph exists for
the set of minima of any such submodular function.  These minima form
a family closed under union and intersection, and any such family
corresponds to the ideals of an appropriate digraph~\cite[Proposition
10.3.3]{Grotschel:1988:GAC}.  Such a representation can be computed efficiently, as
shown by Gabow~\cite{Gabow:1991:APR}.

Intuitively, what Theorem~\ref{theorem-f-to-graph} says is the
following.  For any $\f$, the residues in fittest protein sequences
are grouped into clusters, where the cluster $\rho^{-1}(s)$ is always
$H$, the cluster $\rho^{-1}(t)$ is always $P$, and for each of the
remaining clusters, all residues in the cluster are either all $H$ or
all $P$.  In addition, there is a dependence given by the edges of
$\Gfst$, such that if a cluster corresponding to the source of an edge
is all $H$ then the cluster at the other end is also all $H$.

There is no additional restriction on the structure of the space of
all fittest protein sequences beyond those that follow from correspondence
with the ideals of some digraph.  
As shown in Theorem~\ref{theorem-graph-to-f},
any graph may appear as $\Gfst$, with any number of residues mapped to
each supernode.

\begin{theorem}[characterizing a dag via $\mathbf{\f}$]
\label{theorem-graph-to-f}
Let $\hat{G}$ be an arbitrary digraph with $n$ nodes, labeled $1$ to
$n$, and $m$ edges.  Let $\hat{G}_0$ be the \buzz{component graph} of
$\hat{G}$ obtained by contracting each strongly connected component of
$\hat{G}$ to a single supernode through a contraction map $\kappa$.
Then, there exists some $\f$ as defined in Assumption~\ref{f2} such
that for the $\Gfsta$ and $\rho$ defined in
Theorem~\ref{theorem-f-to-graph}, an isomorphism exists between
$\Gfst$ and $\hat{G}_0$ mapping each $\rho(i)$ to $\kappa(i)$.
\end{theorem}
\begin{proof}
Represent each node $i$ in $\hat{G}$ by
the variable $x_i$.  
For each $i,j$, let $e_{i,j} = 1$ if there is a directed edge
$(i,j)$ in $\hat{G}$, and $0$ otherwise.
To define $\f$,
let $a_{i,j} = e_{i,j} + e_{j,i}$ for each $i<j$;
and, for each $i$, let $b_{i}$ equal $i$'s out-degree 
$\delta^+(i) = \sum_j e_{i,j}$.
Apply
Lemma~\ref{lemma-cuts} to the resulting function $\f$ to get a graph
$G^{\f}$.
Define a flow $\phi$ in $G^{\f}$ as follows:
\begin{displaymath}
\begin{array}{rcll}
\phi(s,u_{i,j}) & = & e_{i,j}+e_{j,i} & \forall u_{i,j} \\
\phi(u_{i,j},v_i) & = & e_{i,j} & \forall u_{i,j} \\
\phi(u_{i,j},v_j) & = & e_{j,i} & \forall u_{i,j} \\
\phi(v_i, t) & = & \delta^+(i) & \forall v_i \\
\end{array}
\end{displaymath}
Note that this flow is a maximum flow because it saturates all edges
leaving $s$ as well as all edges entering $t$.  (It happens that this
is the unique maximum flow, but we do not need this fact, as the
Picard-Queyranne construction works for any maximum flow.)

Our next goal is to show that the residual graph $G^{\f}_\phi$ of this
flow contracts to $\hat{G}_0$.  $G^{\f}_\phi$ has the following
classes of edges:
\begin{center}
\begin{tabular}{|cl|}
\hline
$(u_{i,j},s)$ & \\
$(u_{i,j},v_i)$ & $\forall u_{i,j}$ \\
$(u_{i,j},v_j)$ & \\
\hline
$(v_i,u_{i,j})$ & when $(i,j) \in \hat{G}$\\
\hline
$(v_j,u_{i,j})$ & when $(j,i) \in \hat{G}$\\
\hline
$(v_i,t)$ & $\forall v_i$\\
\hline
\end{tabular}
\end{center}
Since $s$ has no successors and $t$ has no predecessors in
$G^{\f}_\phi$, the supernodes in $\Gfsta$ containing $s$ and $t$
consist of only $s$ and $t$, respectively.  Each node $u_{i,j}$ is in
the same strongly-connected component as at least one of $v_i$ or
$v_j$, so no other supernodes exist in $\Gfsta$ that do not contain at
least one of the nodes $v_i$.  Note that every node-simple path from
$v_i$ to $v_j$ in $G^\f_\phi$ is of the form
$v_i,u_{i,q_1},v_{q_1},u_{q_1,q_2},\ldots,u_{q_{k-1},j},v_j$ (with the
subscripts of each $u_{q_i,q_{i+1}}$ possibly reversed), which
corresponds to a node-simple path $i = q_1,q_2,\ldots,q_k = j$ in
$\hat{G}$.  The converse also holds.  Therefore, $i$ and $j$ are in
the same strongly connected component in $\hat{G}$ if only if $v_i$
and $v_j$ are in the same strongly connected component in $\Gfsta$.
Now recall that for each $i$, $\rho(i)$ is defined in
Lemma~\ref{lemma-cuts} as the supernode in $\Gfsta$ containing $v_i$.
So an edge from $\rho(i)$ to $\rho(j)$ in $\Gfsta$
corresponds to a node-simple path from $v_i$ to $v_j$ $\Gfsta$.  By
the above path-to-path correspondence, every directed edge from
$\rho(i)$ to $\rho(j)$ in $\Gfsta$ corresponds to an edge from
$\kappa(i)$ to $\kappa(j)$, and vice versa.  In summary, $\Gfst$ is
isomorphic to $\hat{G}_0$, with the correspondence $\rho(i)
\leftrightarrow \kappa(i)$ for all $i$.
\end{proof}

\section{Further Tools for Protein Landscape Analysis}
\label{section-further-tools}
\subsection{Optimizing Multiple Objectives}
\label{section-multiobjective}
We can extend the results of Section~\ref{section-basic-tools} beyond
optimizing a single fitness function.  

With more than one fittest protein sequence to choose from, we may
wish to find a fittest protein sequence $x$ that is the closet to some
target protein sequence $\hat{x}$ in unweighted or weighted Hamming
distance.  Theorem~\ref{thm_dist_Hcount} shows that this optimization
problem is as easy as finding an arbitrary fittest protein sequence.

We may also wish to consider what protein sequences are simultaneously
the fittest for more than one fitness function.
Theorem~\ref{theorem-intersection} shows how to compute a
representation of this set similar to that provided by
Theorem~\ref{theorem-f-to-graph}.

\begin{theorem}[optimizing Hamming distances and $H$-residue counts over
$\mathbf{\mopt{\f}}$]
\label{thm_dist_Hcount} 
\ftwo{}
\begin{enumerate}
\item\label{item_theorem-Hamming-distance} Given a target protein
sequence $\hat{x}$, some weights $w_i$, and $\f$ as the input, we can
find in $O(\minCutTime)$ time an $x \in \opt{\f}$ with the minimum
weighted Hamming distance $\sum_i w_i |x_i - \hat{x}_i|$ over
$\opt{\f}$.
\item\label{item_corollary-most-least-H} Given $\f$ as the input, we
can find in $O(\minCutTime)$ time an $x \in \opt{\f}$ with the largest
$($or smallest$)$ possible number of $H$ residues over $\opt{\f}$.
\end{enumerate}
\end{theorem}
\begin{proof} The statements are proved as follows.

Statement~\ref{item_theorem-Hamming-distance}. Let $\epsilon$ be a
positive constant at most $\frac{1}{4Wnc}$, where $W \ge \max |w_i|$
and $c$ is the common denominator of all coefficients $a_{i,j}$ and
$b_i$.  Below we show how to find a desired fittest protein sequence
by minimizing $\f_\epsilon(x)=\f(x)+\sum_i \epsilon
w_i\left|x_i-\hat{x}_i\right|$.

First of all, since $x$ and $\hat{x}$ are 0-1 sequences,
$\left|x_i-\hat{x}_i\right|=(x_i-\hat{x}_i)^2=
x_i-2\hat{x_i}x_i+\hat{x}_i$. Then, since $\hat{x}$ is given,
$\f_\epsilon(x)$ can be minimized using
Lemma~\ref{lemma-minimize-cost} in $O(\minCutTime)$ time.

Now suppose that $y$ and $z$ are two protein sequences with $y \in
\opt{\f}$ and $z \not\in \opt{\f}$. Then, $\f(z)-\f(y) \geq \frac{1}{c}$. 
Also, $\left|\sum_i w_i |x_i-\hat{x}_i|\right| \leq Wn \leq \frac{1}{4c}$.
Therefore,
$\f_\epsilon(z)-\f_\epsilon(y)\geq\frac{1}{c}-2Wn\geq\frac{1}{2c}$. Thus
every $x$ that minimizes $\f_\epsilon(x)$ must also minimize $\f(x)$.
The Hamming distance term in $\f_\epsilon$ guarantees that from all
$x$ that do minimize $\f(x)$, minimizing $\f_\epsilon(x)$ selects one
that also minimizes this distance.

Statement \ref{item_corollary-most-least-H}.  To find an
$x\in\opt{\f}$ with the largest (respectively, smallest) possible
number of $H$ residues, apply
Statement~\ref{item_theorem-Hamming-distance} with all $w_i=1$ and all
$\hat{x}_i=1$ (respectively, $\hat{x}_i= 0$).
\end{proof}

Suppose we are given fitness functions $\f^1,\ldots,\f^f$
corresponding to multiple 3D structures, and we wish to find protein
sequences that are simultaneously optimal for each 3D structure.  A
simple approach is to observe that the function $\f^* = \f^1+\cdots+\f^f$ 
satisfies Assumption~\ref{f2}, and that any protein
sequence that simultaneously optimizes each $\f^\ell$ optimizes
$\f^*$.  However, we must check any minimum solution for $\f^*$ to see
that it is in fact a minimum solution for each $\f^\ell$, as it may be
that the sets of minimum solutions of the $\f^\ell$ have empty
intersection.  Performing both the optimization of $\f^*$ and of the
$f$ individual fitness functions requires $f+1$ network flow
computations.

It turns out that we can reduce this cost to $f$ network flows at the
cost of some additional work to compute a composite Picard-Queyranne
graph $\Gfmultia$ directly from the individual graphs $\Gfstla$.  This
approach, described in Theorem~\ref{theorem-intersection}, is
especially 
useful if we have already computed the individual graphs for some
other purpose.

\begin{theorem}[minimizing multiple fitness functions]
\label{theorem-intersection}
\fthr{} For each $\ell$, let $\Gfstla$ and $\rho^\ell$ be the dag and
map computed from $\f^\ell$ in Theorem~\ref{theorem-f-to-graph}.
Given all $\Gfstla$ and $\rho^\ell$ as the input, there is an
$O(\Delta)$-time algorithm that either $($a$)$ determines that there
is no protein sequence $x$ that simultaneously minimizes $\f^1$
through $\f^f$, or $($b$)$ constructs a dag $\Gfmultia$ with
designated nodes $s'$ and $t'$ and a mapping $\rho^*$ from $\{1,\ldots,n\}$ 
to $V(\Gfmultia)$, such that there is a one-to-one correspondence
between the protein sequences $x$ that simultaneously minimize all
$\f^\ell(x)$ and the ideals of $\Gfmulti = \Gfmultia - \{s',t'\}$, in
which $x_i = 0$ if and only if $\rho^*(i) = t'$ or $\rho^*(i)$ is in
the ideal corresponding to $x$.
\end{theorem}
\begin{proof}
By Theorem~\ref{theorem-f-to-graph}, the conditions below are
necessary and sufficient for $x$ to minimize $\f^\ell$:
\begin{itemize}
\item $x_i = 1$ if $\rho^\ell(i) = s'$.
\item $x_i = 0$ if $\rho^\ell(i) = t'$.
\item $x_i = x_j$ if $\rho^\ell(i) = \rho^\ell(j)$.
\item $x_i \le x_j$ if $(\rho^\ell(i), \rho^\ell(j))$ is an edge in
$\Gfstl$.
\end{itemize}

We will build a graph $G$ whose nodes are $s$, $t$, and
$1,\ldots,n$, and put in an edge $(u,v)$ between any nodes for
which the constraint $u \le v$ is required to minimize some $\f^\ell$.
In particular, we have the following classes of edges, for each
$\ell$, where each class
represents one of the above conditions:
\begin{itemize}
\item $(s,v)$ and $(v,s)$ whenever $\rho^\ell(v) = s'$.
\item $(t,v)$ and $(v,t)$ whenever $\rho^\ell(v) = t'$.
\item $(u,v)$ and $(v,u)$ whenever $\rho^\ell(u) = \rho^\ell(v)$.
\item $(u,v)$ whenever $(\rho^\ell(u),\rho^\ell(v) \in E(\Gfstl))$.
\end{itemize}

An assignment of $1$ to $s$, $0$ to $t$, and $x_i$ to each node $i$
satisfies $u \le v$ whenever $(u,v) \in E(G)$ if and only if all of
the constraints required for $x$ to simultaneously minimize all
$\f^\ell$ are satisfied.  Note that such an assignment might not
exist.

To convert $G$ into the desired graph $\Gfmultia$, and to check
whether there exist any assignments meeting the constraints, contract
each strongly connected component of $G$.  If $s$ and $t$ are in the
same strongly connected component, no simultaneous fittest solutions
exist.  Otherwise, let $s'$ in $\Gfmultia$ be the supernode that
contains $s$ from $G$; let $t'$ be the supernode that contains $t$.
Also, for each $u$ in $1,\ldots,n$, let $\rho^*(u)$ be the supernode
into which $u$ is contracted.  Then $x$ simultaneously minimizes all
$\f^\ell$ if and only if $x_i = 1$ when $\rho^*(i) = s'$, $x_i = 0$
when $\rho^*(i) = t'$, and $x_i \le x_j$ when $(\rho^*(i),\rho^*(j))$
is an edge in $\Gfmultia$---precisely the condition that the zeroes
in $x$ correspond to nodes in some ideal of $\Gfmultia$ that contains
$t'$ but not $s'$.

To show the running time, observe that constructing the graph $G$
takes $O(\Delta)$ time, which dominates the contraction step.
\end{proof}

\subsection{The Space of All Fittest Protein Sequences}
\label{section-optimal}

This section discusses some applications of the representation of the
space $\opt{\f}$ given by Theorem~\ref{theorem-f-to-graph}.
Theorem~\ref{theorem-enumerating-optimal} gives an algorithm to
enumerate this space.  Theorem~\ref{theorem-diameter} gives an
algorithm to compute the diameter of the space in nonnegatively
weighted Hamming distance.  Theorem~\ref{theorem-connectivity} gives
an algorithm to determine connectivity properties of the space with
respect to various classes of mutations.

\begin{theorem}[enumerating all protein sequences] 
\label{theorem-enumerating-optimal}
\ftwo{} \givenGfst{} as the input, the protein sequences in $\opt{\f}$
can be enumerated in $O(n)$ time per protein sequence.
\end{theorem}
\ifabs{}{\begin{proof}
An algorithm of Steiner~\cite{Steiner:1986:AGI} enumerates the
ideals of $\Gfst$ in time $O(|V(\Gfst|) = O(n)$ per ideal.
For each ideal, invert the mapping $\rho$ (in $O(n)$ time)
to recover the corresponding protein sequence $x$.
\end{proof}}

\begin{theorem}[computing the diameter]
\label{theorem-diameter}
\ftwo{} \givenGfst{} as the input, it takes $O(n)$ time to compute the
diameter of $\opt{\f}$ in weighted Hamming distance where the weights
$w_i$ are all nonnegative.
\end{theorem}
\ifabs{}{\begin{proof} 
Any two fittest protein sequences $x$ and $y$ can differ only at
indices $i$ where $\rho(i) \notin \{s,t\}$.  Let $d$ be the total
weight of indices $i \in \rho^{-1}(V(\Gfst))$. Then, $d$ is an upper
bound on the diameter.  It is also a lower bound, as $\emptyset$ and
$V(\Gfst)$ are both ideals of $\Gfst$, and these ideals correspond to
two protein sequences at distance $d$ from each other.
\end{proof}}

We can use $\Gfst$ to determine whether $\opt{\f}$ is connected for
various models of mutations.  For instance, we can determine whether
the space is connected for one-point mutations, in which at most one
residue changes with each mutation and all intermediate protein
sequences must remain the fittest.  More generally, we can determine
the minimum $k$ so that the space is connected where each mutation
modifies at most $k$ residues.

We adopt a general model proposed by Kleinberg \cite{Kleinberg:1999:EAP}.
In the model, there is a system $\Lambda$ of subsets of $\{1,\ldots,n\}$ 
that is {\it closed downward}, i.e., if $A \subseteq B \in
\Lambda$, then $A\in\Lambda$.  Two protein sequences $x$ and $y$ are
\buzz{$\Lambda$-adjacent} if 
they are in $\opt{\f}$ and differ exactly at the positions indexed by
elements of some member of $\Lambda$.  A
\buzz{$\Lambda$-chain} is a sequence of protein sequences in $\opt{\f}$ 
where each adjacent pair is $\Lambda$-adjacent.  Two protein sequences
$x$ and $y$ are \buzz{$\Lambda$-connected} if there exists a
$\Lambda$-chain between $x$ and $y$.  A set of protein sequences is
{\it $\Lambda$-connected} if every pair of elements of the set are
{\it $\Lambda$-connected}.  We would like to tell for any given
$\Lambda$ and $\f$ whether particular protein sequences are
$\Lambda$-connected and whether the entire $\opt{\f}$ is
$\Lambda$-connected.

Kleinberg~\cite{Kleinberg:1999:EAP} gives polynomial-time algorithms for
these problems that take $\Lambda$ as input (via oracle calls) and
depend only on the fact that $\f$ is submodular.  We describe a much
simpler algorithm that uses $\Gfst$ from
Theorem~\ref{theorem-f-to-graph}.  This algorithm not only determines
whether two protein sequences (alternatively, all protein sequences in $\opt{\f}$) are
connected for any given $\Lambda$, but also determines the unique
minimum $\Lambda$ for which the desired connectivity holds.  Almost
all of the work is done in the computation of $\Gfst$; once we have
this representation, we can read off the connectivity of $\opt{\f}$
directly.

\begin{theorem}[connectivity via mutations]
\label{theorem-connectivity}
\ftwo{} The following problems can both be solved in $O(n)$ time.
\begin{enumerate}
\item 
\givenGfst{} and two protein sequences $x$ and $x'$ in $\opt{\f}$ as the input,
compute the maximal elements of the smallest downward-closed set
system $\Lambda$ such that $x$ and $x'$ are $\Lambda$-connected.
\item 
\givenGfst{} as the input,
compute the maximal elements of the smallest downward-closed set
system $\Lambda$ such that $\opt{\f}$ is $\Lambda$-connected.
\end{enumerate}
\end{theorem}
\ifabs{}{\begin{proof} The statements are proved as follows.

Statement 1.  Let $I,I'$ be the ideals in $\Gfst$ such that the sets
of zeros in $x,x'$ are $\rho^{-1}(I),\rho^{-1}(I')$, respectively.
Let the maximal elements of $\Lambda$ be the sets $\rho^{-1}(v)$ over
all $v \in I \symdiff I'$, where $\symdiff$ is the symmetric
difference operator.  Thus, $\Lambda$ consists of these sets and all
of their subsets.  We will show that $\Lambda$ is the smallest
downward-closed set system such that there is a $\Lambda$-chain
between $x$ and $x'$ in $\opt{\f}$.

First, consider some set system $\Lambda'$ where for some $v \in I
\symdiff I'$, $A = \rho^{-1}(v)$ is not in $\Lambda'$.  Recall that
$x$ must be constant at the positions indexed by elements of $A$,
where the constant depends on whether or not $v$ is in the ideal in
$\Gfst$ corresponding to $x$.  Partition $\opt{\f}$ into sets
$\Omega^0$ and $\Omega^1$, where $\Omega^j$ consists of all $z$ with
$z_i = j$ for all $i \in A$.  Since $x$ and $x'$ differ on
$\rho^{-1}(v)$, one of them is in $\Omega^0$ and the other is in
$\Omega^1$.  However, since $A \notin \Lambda'$, no protein sequence
in $\Omega^0$ is $\Lambda'$-adjacent to one in $\Omega^1$. So there is
no $\Lambda'$-chain between $x$ and $x'$.

Conversely, to exhibit a $\Lambda$-chain from $x$ to $x'$, it suffices
to show by iterations a $\Lambda$-chain from $x$ to the protein
sequence $y$ whose zeroes are given by $\rho^{-1}(I \cap I')$; the
case of $x'$ is symmetric.  Let $I_0 = I$.  If at any iteration $I_i =
I \cap I'$, we are done.  Otherwise, let $v$ be a maximal element in
$I_i - (I \cap I')$. Then, $I_{i+1} = I_i-\{v\}$ is also an ideal.
Since $v \in I \symdiff I'$, $\rho^{-1}(v)\in\Lambda$ and the protein
sequences corresponding to $I_i$ and $I_{i+1}$ are $\Lambda$-adjacent.
After at most $n$ such iterations, we reach $y$.

Statement 2.  Let $x$ and $x'$ be the protein sequences in $\opt{\f}$
with the largest and the smallest possible numbers of $H$ residues,
respectively. In other words, $x$ and $x'$ correspond to $\Gfst$ and
its empty ideal, respectively.  If $\Lambda$ includes $\rho^{-1}(v)$
for all $v \in V(\Gfst)$, then by Statement 1, there are
$\Lambda$-chains between any $y \in \opt{\f}$ and $x'$ and thus between
any two protein sequences in $\opt{\f}$. If it does not, then there is
no $\Lambda$-chain between $x$ and $x'$.  In summary, the maximal
elements of $\Lambda$ are the sets $\rho^{-1}(v)$ over all $v \in
V(\Gfst)$.
\end{proof}}

\subsection{Generating Near-Fittest Protein Sequences}
\label{section-suboptimal}

Finding good protein sequences other than the fittest is trickier, as
Lemma~\ref{lemma-LP} breaks down if we are not looking at the fittest
protein sequences.  This section gives two algorithms that avoid this
problem.  Theorem~\ref{theorem-enumerating-suboptimal} describes an
algorithm to generate all protein sequences $x$ in order of increasing
$\f(x)$.  Theorem~\ref{theorem-enumerating-by-Hamming-distance}
describes an algorithm to generate the fittest protein sequences at
different unweighted Hamming distances, which is useful for examining
the trade-off between fitness and distance.

The algorithm for generating all protein sequences $x$ in increasing
order by $\f(x)$ is based on Lemma~\ref{lemma-minimize-cost} and a
general technique for enumerating suboptimal solutions to
combinatorial optimization problems due to Lawler~\cite{Lawler:1972:PCB}.
It is similar to an algorithm of Vazirani and
Yannakakis~\cite{Vazirani:1992:SCT} for enumerating suboptimal cuts.  We
cannot use the Vazirani-Yannakakis algorithm directly because
suboptimal cuts in $G^{\f}$ might include cuts corresponding to
assignments in which $y_{i,j}$ is not equal to $x_i x_j$ for some
$i,j$.

\begin{theorem}[enumerating all protein sequences]
\label{theorem-enumerating-suboptimal}
\ftwo{} With $\f$ as the input, we can enumerate all protein sequences
$x$ in order of increasing $\f(x)$ in time $O(n\minCutTime)$ per
protein sequence.
\end{theorem}
\begin{proof}
For any length-$k$ $0$-$1$ sequence $y=y_1,y_2,\ldots,y_k$, let $A_y$
be the set of all length-$n$ $0$-$1$ sequences $x$ with $x_i = y_i$
for each $y_i$ with $1 \le i \le k$.  Let $\varepsilon$ be the empty
sequence.  Then, $A_\varepsilon$ is the set of all length-$n$
sequences.  Observe that we can find an element $z \in A_y$ that
minimizes $\f(z)$ over $A_y$ in time $O(\minCutTime)$ by setting $z_i
= y_i$ in $\f(z)$ for each $y_i$ and applying
Lemma~\ref{lemma-minimize-cost}.  Furthermore, the set $A_y - \{z\}$
is the disjoint union of the sets $A_{r_i}$ for $k+1
\le i \le n$, where $r_i=z_1,z_2,\ldots,z_{i-1},(1-z_i)$.

To enumerate $x$ in order of increasing $\f(x)$, we maintain a data
structure that represents all protein sequences less those already
returned as a disjoint union of sets of the form $A_y$, together with
an $\f$-minimizing element $z$ for each, organized as a priority queue
with key $\f(z)$.  Initially, the queue contains only
$(A_\varepsilon,z)$, where $z \in \opt{\f}$ is computed using
Lemma~\ref{lemma-minimize-cost} in time $O(\minCutTime)$.  At each
step, the smallest pair $(A_y,z)$ is removed from the priority queue
and is replaced by up to $n$ pairs $(A_{r_i},p_i)$, where $p_i$ is an
$\f$-minimizing element of $A_{r_i}$; $z$ is then returned.  Each such
step requires no more than $n$ applications of
Lemma~\ref{lemma-minimize-cost}, and the cost of the at most $n+1$
priority queue operations is at most $O(n \log(2^n)) = O(n^2)$, giving
a total cost of $O(n\minCutTime)$ per value returned.
\end{proof}

Let $\hat{x}$ be a target protein sequence.  For 
$d\in\{0,\ldots,n\}$, let $\FFT(d)$ be the smallest $\f(x)$ over all
protein sequences $x$ at unweighted Hamming distance $d$ from
$\hat{x}$.  A basic task of landscape analysis is to plot the graph of
$\FFT$. As
Theorem~\ref{theorem-bounded-distance-is-hard}(\ref{lem_dist_d}) in
Section~\ref{section-hard-problems} shows, this task is
computationally difficult in general. Therefore, one way to plot the
graph of $\FFT$ would be to use
Theorem~\ref{theorem-enumerating-suboptimal} to enumerate all protein
sequences $x$ in order of increasing $\f(x)$ until for each $d$, at
least one protein sequence at distance $d$ from $\hat{x}$ has been
enumerated.  This solution may require processing exponentially many
protein sequences before $\FFT$ is fully plotted. As an alternative,
Theorem~\ref{theorem-enumerating-by-Hamming-distance} gives a tool for
plotting $\FFT$ approximately in polynomial time.

\begin{theorem}[approximately plotting the energy-distance landscape]
\label{theorem-enumerating-by-Hamming-distance}
\ftwo{} For each $\epsilon$, let $\f_\epsilon(x) = \f(x) +
\epsilon{\cdot}|x-\hat{x}|$.  Let $\FF(\epsilon)$ be the minimum
$\f_\epsilon(x)$ over all $x$.
\begin{enumerate}
\item\label{thm_ed_1}
$\FF$ is a continuous piecewise linear concave function defined on
$\R$ with at most $n+1$ segments and thus at most $n+1$ corners.
\item\label{thm_ed_2}
Let $(\epsilon_1,\FF(\epsilon_1)),\ldots,(\epsilon_k,\FF(\epsilon_k))$
be the corners of $\FF$, where $\epsilon_1 < \cdots < \epsilon_k$.
Let $d_i$ be the slope of the segment immediately to the right of
$\epsilon_i$.  Let $d_0$ be the slope of the segment immediately to
the left of $\epsilon_1$. Then, $n=d_0>d_1>\cdots>d_k=0$.
\item\label{thm_ed_4}
Let $d \in \{0,1,\ldots,n\}$.
\begin{enumerate}
\item\label{thm_ed_4a} 
$\FFT(d_0) = \FF(\epsilon_1)-\epsilon_1{\cdot}d_0$.
$\FFT(d_k) = \FF(\epsilon_k)-\epsilon_k{\cdot}d_k$.
For $0 < i < k$, $\FFT(d_i) = \FF(\epsilon_i)-\epsilon_i{\cdot}d_i
= \FF(\epsilon_{i+1})-\epsilon_{i+1}{\cdot}d_i$.
\item\label{thm_ed_4c} 
For $d_{i}>d>d_{{i+1}}$ with $0 \leq i < k$, 
$\FFT(d)\geq\lambda\FFT(d_i)+(1-\lambda)\FFT(d_{i+1})$, where
$\lambda=\frac{d-d_{i+1}}{d_i-d_{i+1}}$.
\end{enumerate}
\item\label{thm_ed_3}
Given $\f$ and $\hat{x}$ as the input, we can compute
$(\epsilon_1,\FF(\epsilon_1)),\ldots,(\epsilon_k,\FF(\epsilon_k))$ and
$d_{0},\ldots,d_{k}$ in $O(n\minCutTime)$ time.
\end{enumerate}
\end{theorem}
\begin{proof} The statements are proved as follows.

Statement \ref{thm_ed_1}.  The concavity follows from the minimality
of $\FF(\epsilon)$ and the fact that for any fixed $x$,
$\f_\epsilon(x)$ is linear in $\epsilon$ with slope $|x-\hat{x}|$.
Then, the continuous piecewise linearity and the counts of segments
and corners follow from the fact that $|x-\hat{x}| \in
\{0,1,\ldots,n\}$.

Statement \ref{thm_ed_2}. By the concavity of $\FF$,
$d_0>d_1>\cdots>d_k$. 
Let $W=1+\sum_{1\leq i < j \leq n}a_{i,j}+\sum_{i=1}^n|s_i|$. 
For all $\epsilon \leq -W$,
$\f_\epsilon(x)$ is minimized if and only if $x$ is at distance $n$
from $\hat{x}$. Similarly, for all $\epsilon
\geq W$, $\f_\epsilon(x)$ is minimized if and only if $x$ is at
distance $0$ from $\hat{x}$. Therefore, $d_0=n$ and $d_k=0$.

Statement \ref{thm_ed_4}.  Case~\ref{thm_ed_4a} is straightforward.
To prove Case~\ref{thm_ed_4c}, let $x$ be a protein sequence that has
the smallest $\f(x)$ over all protein sequences at distance $d$ from
$\hat{x}$. Then, $\FFT(d) = \f(x)$, and
$\f_{\epsilon_{i+1}}(x)=\FFT(d)+\epsilon_{i+1}{\cdot}d$.  On the other
hand, by the minimality of $\FF$, $\f_{\epsilon_{i+1}}(x) \geq
\FF(\epsilon_{i+1})$.
Furthermore, by Case~\ref{thm_ed_4a}, $\FF(\epsilon_{i+1})
=\FFT(d_{i+1})+\epsilon_{i+1}{\cdot}d_{i+1}
=\FFT(d_i)+\epsilon_{i+1}{\cdot}d_i$. Thus,
\begin{displaymath}
\begin{array}{rcl}
\FFT(d_{i+1}) & \leq & \FFT(d)+\epsilon_{i+1}{\cdot}d-\epsilon_{i+1}{\cdot}d_{i+1};
\\
\FFT(d_i) & \leq & \FFT(d)+\epsilon_{i+1}{\cdot}d-\epsilon_{i+1}{\cdot}d_i.
\end{array}
\end{displaymath}
Case~\ref{thm_ed_4c} follows from algebra and these two inequalities.

Statement \ref{thm_ed_3}.  For given $\epsilon$ and $x$, let
$\LL(\epsilon,x)$ be the line through the point
$(\epsilon,\f_{\epsilon}(x))$ and with slope $|x-\hat{x}|$.  Let
$L_\epsilon$ (respectively, $R_\epsilon$) be the protein sequence $x$
such that $|x-\hat{x}|$ is the largest (respectively, smallest)
possible over $\opt{\f_\epsilon}$.  Note that $L_\epsilon$,
$R_\epsilon$, $\LL(\epsilon,L_\epsilon)$, and
$\LL(\epsilon,R_\epsilon)$ can be computed in $O(\minCutTime)$ total
time using Theorem~\ref{thm_dist_Hcount}.  Furthermore,
$\LL(\epsilon,L_\epsilon)$ and $\LL(\epsilon,R_\epsilon)$ contain the
segments of $\FF$ immediately to the left and the right of $\epsilon$,
respectively. Consequently, $\epsilon$ is a corner of $\FF$ if and
only if $\LL(\epsilon,L_\epsilon)\not=\LL(\epsilon,R_\epsilon)$.

To compute the corners and slopes of $\FF$, we first describe a
recursive corner-slope finding subroutine as follows.  The subroutine
takes as input an interval $[\epsilon',\epsilon'']$ where $\epsilon' <
\epsilon''$ together with $\LL(\epsilon',R_{\epsilon'})$ and
$\LL(\epsilon'',L_{\epsilon''})$. It outputs all the corners
$(\epsilon,\FF(\epsilon))$ of $\FF$ together with slopes
$|L_\epsilon-\hat{x}|$ and $|R_\epsilon-\hat{x}|$ where
$\epsilon'<\epsilon<\epsilon''$. There are two cases.

{\it Case} 1:
$\LL(\epsilon',R_{\epsilon'})=\LL(\epsilon'',L_{\epsilon''})$.  Then,
there is no corner over the interval $(\epsilon',\epsilon'')$, and
thus the subroutine call ends without reporting any new corner or
slope.

{\it Case} 2:
$\LL(\epsilon',R_{\epsilon'})\not=\LL(\epsilon'',L_{\epsilon''})$.
Then, compute $\epsilon'''$ at which $\LL(\epsilon',R_{\epsilon'})$
and $\LL(\epsilon'',L_{\epsilon''})$ intersect; by the concavity of
$\FF$ stated in Statement~\ref{thm_ed_1}, $\epsilon' < \epsilon''' <
\epsilon''$.  Also, compute $\LL(\epsilon''',L_{\epsilon'''})$ and
$\LL(\epsilon''',R_{\epsilon'''})$. There are two subcases:

{\it Case} 2a:
$\LL(\epsilon''',L_{\epsilon'''})\not=\LL(\epsilon''',R_{\epsilon'''})$.
Then the subroutine returns $(\epsilon''',\FF(\epsilon'''))$ as a new
corner together with slopes $|L_{\epsilon'''}-\hat{x}|$ and
$|R_{\epsilon'''}-\hat{x}|$ and recurses on the intervals
$[\epsilon',\epsilon''']$ and $[\epsilon''',\epsilon'']$.

{\it Case} 2b: $\LL(\epsilon''',L_{\epsilon'''})=\LL(\epsilon''',R_{\epsilon'''})$.
The subroutine returns no new corner or slope but recurses on the intervals
$[\epsilon',\epsilon''']$ and $[\epsilon''',\epsilon'']$. In this case, the
subroutine has found the line containing a new segment of $\FF$, i.e.,
the segment through the point $(\epsilon''',\FF(\epsilon'''))$.

This completes the description of the subroutine.  The running time of
this subroutine is dominated by that for computing $L_{\epsilon'''}$ and
$R_{\epsilon'''}$ and thus is $O(\minCutTime)$.

With this subroutine, we can find the corners and slopes of $\FF$ as
follows.  Recall $W$ from the proof of Statement \ref{thm_ed_2}.  Note
that if $\epsilon \leq -W$ or $\epsilon \geq W$, then $\FF$ has no
corner at $\epsilon$. So we compute $\LL(-2W,R_{-2W})$ and
$\LL(2W,L_{2W})$ and apply the subroutine to the interval $[-2W,2W]$
to find all the corners and slopes of $\FF$.  This algorithm makes
$O(n)$ recursive calls to the subroutine since by
Statement~\ref{thm_ed_1}, there are only $O(n)$ corners and segments,
and each recursive call finds at least one new corner or segment. The
running time of the algorithm is dominated by the total running time
of these calls and thus is $O(n\minCutTime)$ as stated in the
statement.
\end{proof}

\subsection{Tuning the Parameters of the GC Model}\label{section-tuning}
This section shows how to systematically tune the parameters $\alpha$
and $\beta$ so that a fittest protein sequence for a given 3D
structure matches the 3D structure's native protein sequence as
closely as possible in terms of unweighted or weighted Hamming
distance.  For this purpose, we assume $s_i \geq 0$.  Furthermore,
since the fitness function does not have an absolute scale, we may fix
$\alpha$ at $-1$ and vary $\beta$. In summary, this section adopts
Assumption~\ref{f4}.

Let $\Pi_{\f}$ be the set of indices $i$ with $s_i \neq 0$.  The next
lemma shows that for any fittest protein sequence $x$ of $\f_\beta$, the set
$H(x)\cap\Pi_{\f}$ of $H$ residues with nonzero surface area $s_i$ is
monotone in $\beta$.  Then, as shown in Theorem~\ref{theorem-tuning},
to tune $\beta$, we only need to consider at most $n+1$ possible
values of $\beta$.

\begin{lemma}\label{lemma-monotonicity}
\ffou{}
Let $y$ and $z$ be any fittest protein sequences for $\f_{\beta_1}$ and
$\f_{\beta_0}$, respectively.  If $\beta_1 < \beta_0$, then
$H(y)\cap\Pi_{\f} \supseteq H(z)\cap\Pi_{\f}$.
\end{lemma} 
\begin{proof} 
Let $H_1 = H(y)$ and $H_0 = H(z)$.  Let $H_{10} = H_1 - H_0$, i.e., the
set of indices $i$ for which $x_i$ changes from $1$ to $0$ when
$\beta$ changes from $\beta_1$ to $\beta_0$. Similarly, let $H_{01} =
H_0 - H_1$. Further, 
let $\KK_{10}=\sum_{i \in H_{10}} s_i$ and 
    $\KK_{01} = \sum_{i \in H_{01}} s_i$.  

Let $\GG = \sum_{i,j \in H_0} a_{i,j} - \sum_{i,j \in
H_1} a_{i,j}$.  Then, $\f_{\beta}(z)-\f_{\beta}(y)=-\GG+\beta(\KK_{01}-\KK_{10})$.
Let $\GG_{01} = \sum_{i,j \in H_0\ \wedge\ \{i,j\}\cap H_{01}
\not=\emptyset} a_{i,j}$.  Let $\GG_{10} = -(\GG - \GG_{01})$, which
is the sum of the terms $a_{i,j}$ that $\f_{\beta}$ loses when $x_i$
changes from $1$ to $0$.  Similarly, $\GG_{01}$ is the sum of the
terms $a_{i,j}$ that $\f_{\beta}$ gains when $x_i$ changes from $1$ to
$0$.

To show $H(y)\cap\Pi_{\f} \supseteq H(z)\cap\Pi_{\f}$, we need to prove
$H_{01}\cap\Pi_{\f}=\emptyset$ or equivalently $\KK_{01}=0$.  To do so by
contradiction, suppose $\KK_{01} > 0$.  There are two cases:

{\it Case} 1: $-\GG_{01} + \beta_0 \KK_{01} > 0$.  Notice that the
protein sequence $z'$ with $H(z')=H_0-H_{01}$ has a smaller fitness value for
$\beta_0$ than $z$ does, contradicting the minimality of $z$.

{\it Case} 2: $-\GG_{01} + \beta_0 \KK_{01} \leq 0$.  Then, $-\GG_{01}
+ \beta_1 \KK_{01} < 0$. Therefor, the protein sequence $y'$ with $H(y') = H_1
\cup H_{01}$ has a smaller fitness value for $\beta_1$ than $y$ does,
contradicting the minimality of $y$.
\end{proof}

Let $\FF(\beta)$ be the minimum $\f_\beta(x)$ over all $x$.  The next
lemma characterizes the structure of $\FF(\beta)$. This structure is
then used to tune $\beta$ in Theorem~\ref{theorem-tuning}.

\begin{lemma}\label{lem_concavity}
\ffou{}
\begin{enumerate}
\item\label{concavity_1} $\FF$ is a continuous piecewise linear concave function
defined on $[0,\infty)$ with at most $n+1$ segments and thus at most
$n+1$ corners.
\item\label{concavity_3} For all $\beta_1<\beta_3< \beta_4 <\beta_2$ where
$(\beta_1,\FF(\beta_1))$ and $(\beta_2,\FF(\beta_2))$ are adjacent
corners of $\FF$, we have $\opt{\f_{\beta_3}}=\opt{\f_{\beta_4}}$,
$\opt{\f_{\beta_3}}\subseteq\opt{\f_{\beta_1}}$, and
$\opt{\f_{\beta_3}}\subseteq\opt{\f_{\beta_2}}$.  Similarly,
for all $\beta_1<\beta_3<\beta_4$ where $(\beta_1,\FF(\beta_1))$ is the
rightmost corner, we have
$\opt{\f_{\beta_3}}=\opt{\f_{\beta_4}}$
and $\opt{\f_{\beta_3}}\subseteq\opt{\f_{\beta_1}}$.
\item\label{concavity_2} 
Given $\f$ as the input, 
it takes $O(n\minCutTime)$ time to find the set of all $\beta$ such
that $(\beta,\FF(\beta))$ is a corner of $\FF$.
\end{enumerate}
\end{lemma}
\begin{proof} The statements are proved as follows.

Statement~\ref{concavity_1}.  The concavity follows from the
minimality of $\FF(\beta)$ and the fact that for any fixed $x$,
$\f_\beta(x)$ is linear in $\beta$ with slope $\sum_{i \in
H(x)\cap\Pi_\f} s_i$.  Then, the continuous piecewise linearity and the counts of
segments and corners follow from Lemma~\ref{lemma-monotonicity}.

Statement~\ref{concavity_3}.  The proofs for the case that
$(\beta_1,\FF(\beta_1))$ is the rightmost corner and the complementary
case are similar. So we only detail the proof for the former.  Note
that $(\beta_3,\FF(\beta_3))$ is not a corner. So for every fixed
$x\in\opt{\f_{\beta_3}}$, the line $\f_\beta(x)$ goes through the
corner $(\beta_1,\FF(\beta_1))$ and the point
$(\beta_4,\FF(\beta_4))$.  Thus, $x\in\opt{\f_{\beta_1}}$ and
$x\in\opt{\f_{\beta_4}}$.  By symmetry, for every $y \in
\opt{\f_{\beta_4}}$, we have $y\in\opt{\f_{\beta_3}}$. In summary,
$\opt{\f_{\beta_3}}=\opt{\f_{\beta_4}}$ and
$\opt{\f_{\beta_3}}\subseteq\opt{\f_{\beta_1}}$.

Statement~\ref{concavity_2}.  For any given $\beta$ and $x$, let
$\LL(\beta,x)$ be the line through the point $(\beta,\f_{\beta}(x))$
and with slope $\sum_{i \in H(x)\cap\Pi_\f} s_i$.  For each $\beta$,
let $L_\beta$ (respectively, $R_\beta$) be the protein sequence $x$
such that $H(x)$ has the largest (respectively, smallest) possible
cardinality over $\opt{\f_\beta}$.  Note that $L_\beta$, $R_\beta$,
$\LL(\beta,L_\beta)$, and $\LL(\beta,R_\beta)$ can be computed in
$O(\minCutTime)$ total time using Theorem~\ref{thm_dist_Hcount}.
Furthermore, $\LL(\beta,L_\beta)$ and $\LL(\beta,R_\beta)$ contain the
segments of $\FF$ immediately to the left and the right of $\beta$,
respectively. Consequently, for $\beta > 0$, $\beta$ is a corner of
$\FF$ if and only if
$\LL(\beta,L_\beta)\not=\LL(\beta,R_\beta)$. Also, $(0,\FF(0))$ is the
leftmost corner, and the segment of $\FF$ to the right of $0$ is
contained by $\LL(0,R_0)$.

To compute the corners of $\FF$, we first describe a recursive
corner-finding subroutine as follows.  The subroutine takes as input
an interval $[\beta_1,\beta_2]$ where $\beta_1 < \beta_2$ together
with $\LL(\beta_1,R_{\beta_1})$ and $\LL(\beta_2,L_{\beta_2})$. It
outputs all the corners $(\beta,\FF(\beta))$ of $\FF$ with $\beta_1 <
\beta <
\beta_2$. There are two cases.

{\it Case} 1: $\LL(\beta_1,R_{\beta_1})=\LL(\beta_2,L_{\beta_2})$.
Then, there is no corner over the interval $(\beta_1,\beta_2)$, and
thus the subroutine call ends without reporting any new corner.

{\it Case} 2: $\LL(\beta_1,R_{\beta_1})\not=\LL(\beta_2,L_{\beta_2})$.
Then, compute $\beta_3$ at which $\LL(\beta_1,R_{\beta_1})$ and
$\LL(\beta_2,L_{\beta_2})$ intersect; by the concavity of $\FF$ stated
in Lemma~\ref{lem_concavity}(\ref{concavity_1}), $\beta_1 < \beta_3 <
\beta_2$.  Also, compute $\LL(\beta_3,L_{\beta_3})$ and
$\LL(\beta_3,R_{\beta_3})$. There are two subcases:

{\it Case} 2a:
$\LL(\beta_3,L_{\beta_3})\not=\LL(\beta_3,R_{\beta_3})$.  Then the
subroutine returns $(\beta_3,\FF(\beta_3))$ as a new corner and
recurses on the intervals $[\beta_1,\beta_3]$ and $[\beta_3,\beta_2]$.

{\it Case} 2b: $\LL(\beta_3,L_{\beta_3})=\LL(\beta_3,R_{\beta_3})$.
The subroutine returns no new corner but recurses on the intervals
$[\beta_1,\beta_3]$ and $[\beta_3,\beta_2]$. In this case, the
subroutine has found the line containing a new segment of $\FF$, i.e.,
the segment through the point $(\beta_3,\FF(\beta_3))$.

This completes the description of the subroutine.  The running time of
this subroutine is dominated by that for computing $L_{\beta_3}$ and
$R_{\beta_3}$ and thus is $O(\minCutTime)$.

With this subroutine, we can find the corners of $\FF$ as follows.  If
every $s_i = 0$, then $(0,\FF(0))$ is the only corner.  Otherwise, let
$\beta_\infty$ be $1+\sum_{1 \leq i < j \leq n} a_{i,j}$ divided by
the smallest nonzero $s_i$. Note that for every
$\beta\geq\beta_\infty$, $\FF$ has no corner at $\beta$, Then, we
compute $\LL(0,L_0)$ and $\LL(0,R_{\beta_\infty})$. We report the
leftmost corner $(0,\FF(0))$ and apply the subroutine to the interval
$[0,\beta_\infty]$ to find all the other corners of $\FF$.

This algorithm makes $O(n)$ recursive calls to the subroutine since by
Lemma~\ref{lem_concavity}, there are only $O(n)$ corners and segments,
and each recursive call finds at least one new corner or segment. The
running time of the algorithm is dominated by the total running time
of these calls and thus is $O(n\minCutTime)$ as stated in the lemma.
\end{proof}

\begin{theorem}[tuning $\mathbf{\alpha}$ and $\mathbf{\beta}$]
\label{theorem-tuning}
\ffou{}
Given a target protein sequence $\hat{x}$ and $\f$ as the input, we can find
in $O(n\minCutTime)$ time the set of all $\beta$ where the closest
unweighted $($or weighted$)$ Hamming distance between $\hat{x}$ and
any protein sequence in $\opt{\f_\beta}$ is the minimum over all possible
$\beta$.
\end{theorem}
\begin{proof}
The proofs for the cases of unweighted and weighted Hamming distances
are similar.  So we only detail the proof for the unweighted case.
Our algorithm for finding all distance-minimizing choices of $\beta$ has
three stages.

Stage 1.  Use Lemma~\ref{lem_concavity}(\ref{concavity_2}) to find all
the corners of $\FF$.

Stage 2. For each corner $(\beta,\FF(\beta))$, use
Theorem~\ref{thm_dist_Hcount} to compute the closest Hamming distance
$d_\beta$ between $\hat{x}$ and any protein sequence in
$\opt{\f_\beta}$. Let $d_{\rm min}$ be the smallest $d_\beta$ over all
corners. Then, by Lemma~\ref{lem_concavity}(\ref{concavity_3}), report
all $\beta$ with $d_\beta = d_{\rm min}$ as desired choices of
distance-minimizing $\beta$.

Stage 3. Consider each segment of $\FF$. Let $\beta_1$ and $\beta_2$
be the vertical coordinates of the left and right endpoints of the
segment. Find a suitable $\beta_3$ in the open interval
$(\beta_1,\beta_2)$ as follows.  If $\beta_2$ is finite, then set
$\beta_3 = (\beta_1+\beta_2)/2$; otherwise, set $\beta_3 = \beta_1+1$.
Then use Theorem~\ref{thm_dist_Hcount} to compute the closest
unweighted or weighted Hamming distance $d_{\beta_3}$ between
$\hat{x}$ and any protein sequence in $\opt{\f_{\beta_3}}$. If
$d_{\beta_3}=d_{\rm min}$, then by
Lemma~\ref{lem_concavity}(\ref{concavity_3}), report that every
$\beta$ in the interval $(\beta_1,\beta_2)$ is a desired
distance-minimizing $\beta$.

This completes the description of the algorithm.  By
Lemma~\ref{lem_concavity}(\ref{concavity_2}), Stage 1 takes
$O(n\minCutTime)$ time. By Theorem~\ref{thm_dist_Hcount} and
Lemma~\ref{lem_concavity} (\ref{concavity_1}), Stages 2 and 3 also
take $O(n\minCutTime)$ time. Thus, the total running time is as stated
in the theorem.
\end{proof}

\section{Computational Hardness Results}
\label{section-hard-problems}

\begin{theorem}[hardness of counting and averaging]
\label{theorem-counting-optimal-is-hard}
\ftwo{} The following problems are all \sharpP-complete:
\begin{enumerate}
\item\label{theorem-counting-optimal-is-hard_1} Given $\f$ as the
input, compute the cardinality of $\opt{\f}$.

\item\label{corollary-counting-intersection-is-hard} Given
$\f^1,\ldots,\f^f$ as the input, where $f$ is any fixed positive
integer and \fthrr, compute the number of protein sequences $x$ that
simultaneously minimize $\f^\ell(x)$ for all $\ell=1,\ldots,f$.

\item\label{theorem-counting-optimal-is-hard_2} Given $\f$ as the
input, compute the average norm $|x|$, i.e., the average number of $H$
residues in $x$, over all $x\in\opt{\f}$.
\item\label{theorem-counting-optimal-is-hard_3} Given $\f$ and a
target protein sequence $\hat{x}$ as the input, compute the average
unweighted Hamming distance $|x-\hat{x}|$ over all $x\in\opt{\f}$.
\item\label{theorem-counting-optimal-is-hard_4} Given $\f$, a target
protein sequence $\hat{x}$, and an integer $d$ as the input, compute
the number of protein sequences in $\opt{\f}$ at unweighted Hamming
distance $d$ from $\hat{x}$.
\end{enumerate}
\end{theorem}
\ifabs{}{\begin{proof} Note that each of the problems is in \sharpP,
because we can recognize an element of $\opt{\f}$ in polynomial time
using Lemma~\ref{lemma-minimize-cost}.  So to prove
\sharpP-completeness we must only show that each problem is
\sharpP-hard.

Statement~\ref{theorem-counting-optimal-is-hard_1}.  Reduce from the
problem of counting the number of ideals in a dag, which is
\sharpP-hard~\cite{Provan:1983:CCC}.  Given a dag $\hat{G}$, apply
Theorem~\ref{theorem-graph-to-f} to get a function $\f$ for which
$\Gfst$ is isomorphic to $\hat{G}$.  By
Theorem~\ref{theorem-f-to-graph}, counting $\opt{\f}$ is then
equivalent to counting the number of ideals of $\Gfst \isom \hat{G}$.

Statement~\ref{corollary-counting-intersection-is-hard}.  The problem
in Statement~\ref{theorem-counting-optimal-is-hard_1} is a special
case of the problem in this statement.

Statement~\ref{theorem-counting-optimal-is-hard_2}. Using the same
construction as in Statement~\ref{theorem-counting-optimal-is-hard_1},
we can reduce from the problem of computing the average cardinality of
an ideal in $\hat{G}$.  To see that this latter problem is
\sharpP-hard, suppose that we can compute the average cardinalities of
ideals in an $n$-node $\hat{G}$ and in an augmented graph $\hat{G}'$
obtained from $\hat{G}$ by adding a single new node $s$ and edges from
every $v \in V(\hat{G})$ to $s$.  Let $c$ be the average for $\hat{G}$
and $c'$ the average for $\hat{G}'$.  Let $N$ be the number of ideals
in $\hat{G}$.  Then $c = K/N$ for some $K$, while $c' =
(K+n+1)/(N+1)$, since the only new ideal in $\hat{G}'$ consists of $s$
and all other nodes, and thus has size $n+1$.  Solving for $N$ gives
$N = (n+1-c')/(c'-c)$, which can be computed from $c$, $c'$, and $n$.

Statement~\ref{theorem-counting-optimal-is-hard_3}.  This problem has
the problem in Statement~\ref{theorem-counting-optimal-is-hard_2} as a
special case with all $\hat{x_i} = 0$.

Statement~\ref{theorem-counting-optimal-is-hard_4}.  To reduce the
problem of counting protein sequences in $\opt{\f}$ to counting
protein sequences at a given unweighted Hamming distance, take the dag
$\Gfst$ given by Theorem~\ref{theorem-f-to-graph}, and add to each
node $v_i$ a node $q_i$ with edges $(v_i,q_i)$ and $(q_i,v_i)$.  Apply
Theorem~\ref{theorem-graph-to-f} to this new graph $\hat{G}'$ to
obtain a function $\f'$ for which $\opt{\f'}$ is in one-to-one
correspondence with the set of ideals of the strongly-connected
component graph $\hat{G}'_0$ of $\hat{G}'$.  Each strongly connected
component of $\hat{G}'$ consists of $v_i$ and $q_i$ for some $i$, so
$\hat{G}'_0$ is isomorphic to $\Gfst$.  Now we choose $\hat{x}$ with
$\hat{x}_{v_i} = 0$ and $\hat{x}_{q_i} = 1$.  Then, the contribution
to $|x - \hat{x}|$ of each pair $x_{v_i},x_{q_i}$ is $1$ regardless of
their common value.  Thus, the number of $\f$-minimizing protein
sequences equals the number of $\f'$-minimizing protein sequences at
distance $d$ from $\hat{x}$, where $d$ is the number of nodes in
$\Gfst$.
\end{proof}}

\begin{theorem}[hardness of plotting the energy-distance landscape]
\label{theorem-bounded-distance-is-hard}
\ftwo{}
\begin{enumerate}
\item\label{thm_bound_d_1}
Given $\f$ and two integers $d_1,d_2$ as the input, it is
\NP-complete to determine whether there is an $\f$-minimizing $x$ with
$d_1 \le |x| \le d_2$.
\item\label{lem_dist_d} Let $\hat{x}$ be a target protein sequence.
For $d\in\{0,\ldots,n\}$, let $\FFT(d)$ be the smallest $\f(x)$ over
all protein sequences $x$ at unweighted Hamming distance $d$ from
$\hat{x}$.  Given $\f$ and $d$ as the input, it is NP-hard to compute
$\FFT(d)$.
\end{enumerate}
\end{theorem}
\ifabs{}{\begin{proof} Statement~\ref{lem_dist_d} follows from the
fact that the problem in Statement~\ref{thm_bound_d_1} can be reduced
to the problem in this statement in polynomial time. 
Statement~\ref{thm_bound_d_1} is proved as follows.

Since we can recognize $\f$-minimizing protein sequences using
Lemma~\ref{lemma-minimize-cost}, the problem is clearly in \NP.  To
show that it is \NP-hard, we reduce from PARTIALLY ORDERED KNAPSACK,
problem MP12 from Garey and
Johnson~\cite[pp.~247--248]{Garey:1979:CIG}.

The input to PARTIALLY ORDERED KNAPSACK consists of a
partially-ordered set $U$, each element $u$ of which is assigned a
size $s(u) \in \Z^+$ and a value $v(u) \in \Z^+$, together with a
upper bound $B_1$ on total size and a lower bound $B_2$ on total
value.  The problem is to determine whether there exists an ideal $I$
in $U$ such that $\sum_{u\in I} s(u) \le B_1$ and $\sum_{u \in I} v(u)
\ge B_2$.  Garey and Johnson note that the problem, even with $s(u) =
v(u)$ for all $u \in U$, is \NP-complete in the strong sense (meaning
that there is some polynomial bound on the size of all numbers in the
input with which it remains \NP-complete).

Given an instance of PARTIALLY ORDERED KNAPSACK with $s(u) = v(u)$ for
all $u$ and all numbers bounded by some polynomial $p$, build a graph
$\hat{G}$ where each $u \in U$ is represented by a clique $C(u)$ of
$s(u)$ nodes, and there is an edge from $C(u)$ to $C(u')$ if and only if 
$u \prec u'$.  Note that because $s(u) \le p$, $\hat{G}$ has
polynomial size.  Apply Theorem~\ref{theorem-graph-to-f} to generate a
function $\f$ (in polynomial time) such that $\Gfst$ is isomorphic to
the component graph $\hat{G}_0$ obtained by contracting all strongly
connected components of $\hat{G}$.  Since the strongly connected
components of $\hat{G}$ are precisely the cliques $C(u)$, $\Gfst \isom
\hat{G}_0$ is isomorphic to $U$, interpreted as a dag.  In particular
any ideal of $\Gfst$ corresponds to an ideal $I$ of $U$.  Let $N =
\sum_{u \in U} s(u)$.  The norm of the corresponding vector $|x|$ is
$N-\sum_{u \in I} |C(u)| = N-\sum_{u \in I} s(u) = N-\sum_{u \in I}
v(u)$.  Set $d_1 = N-B_1$, $d_2 = N-B_2$, and we have the problem stated
in the theorem.
\end{proof}}

\ifabs{
\acknowledgments
\dobib
\appendix
}{}

\ifabs{
\section*{Appendix: Applications to Empirical Protein 3D Structures}\label{section-pdb}
}{
\section{Applications to Empirical Protein 3D Structures}\label{section-pdb}
}
\begin{table}
\begin{center}
\begin{tabular}{|l|r|r|r|l|l|}
\hline
Name 
& 
Solvent/Length 
&
Length
&
$\alpha/\beta$ 
&
\% Similarity 
&
Description 
\\ \hline\hline
1a7m 
&
51.23 
&
180 
&
-295.8 
&
74.44 
&
cytokine
\\ \hline
1a8y
&
81.37
&
338
&
-155.7
&
73.37
&
Ca binding protein
\\ \hline
1ab3
&
399.05
&
88
&
-326.7
&
78.41
&
ribosomal protein
\\ \hline
1ab7
&
451.48
&
89
&
-79.5
&
80.90
&
ribonuclease inhibitor
\\ \hline
1agi
&
384.30
&
125
&
-93
&
77.60
&
endonuclease
\\ \hline
1air
&
173.96
&
352
&
-0.3
&
69.89
&
pectate lyase
\\ \hline
1b71
&
374.23
&
191
&
-23.7
&
69.11
&
electron transport
\\ \hline
1ble
&
498.45
&
161
&
-269.4
&
72.67
&
phosphotransferase
\\ \hline
1bpi
&
1453.75
&
58
&
-31.5
&
68.97
&
proteinase inhibitor
\\ \hline
1bw3
&
732.31
&
125
&
-33.9
&
70.40
&
lectin
\\ \hline
1clh
&
607.48
&
166
&
-9.3
&
69.28
&
cyclophilin
\\ \hline
1ehs
&
2178.29
&
48
&
-32.1
&
72.92
&
enterotoxin
\\ \hline
1gym
&
392.97
&
296
&
-4.5
&
73.99
&
phospholipase
\\ \hline
1nar
&
447.48
&
289
&
-6.3
&
75.78
&
plant seed protein
\\ \hline
1prn
&
498.38
&
289
&
-2.7
&
56.40
&
porin
\\ \hline
1thv
&
741.41
&
207
&
-10.5
&
71.01
&
sweet tasting protein
\\ \hline
1xnb
&
871.63
&
185
&
-135.6
&
65.95
&
glycosidase
\\ \hline
2aak
&
1130.39
&
150
&
-149.1
&
78.67
&
ubiquitin conjugation
\\ \hline
2bnh
&
412.43
&
456
&
-253.8
&
78.51
&
ribonuclease inhibitor
\\ \hline
2cba
&
773.98
&
258
&
-2.4
&
73.64
&
lyase
\\ \hline
2erl
&
5064.83
&
40
&
-47.7
&
80.00
&
pheromone
\\ \hline
2stv
&
1153.80
&
184
&
-3.6
&
64.67
&
viral coat protein
\\ \hline
6yas
&
904.99
&
256
&
-12.6
&
67.19
&
lyase
\\ \hline
8cho
&
2054.72
&
125
&
-423.6
&
72.80
&
isomerase
\\ \hline
9rat
&
2126.95
&
124
&
-89.4
&
75.00
&
ribonuclease A
\\ \hline\hline
{\bf 1aaj}
&
51.89
&
105
&
-155.1
&
70.48 (72) 
&
electron transport
\\ \hline
{\bf 1aba}
&
125.39
&
87
&
-245.4
&
78.16 (70)
&
electron transport
\\ \hline
{\bf 1bba}
&
584.25
&
36
&
-69.6
&
66.67 (58)
&
pancreatic hormone
\\ \hline
{\bf 1brq}
&
272.54
&
174
&
-27.3
&
72.99 (71)
&
retinol transport
\\ \hline
{\bf 1cis}
&
780.88
&
66
&
-454.8
&
69.70 (64)
&
lysozyme
\\ \hline
{\bf 1hel}
&
539.14
&
129
&
-18.6
&
76.74 (78)
&
fatty acid bind protein
\\ \hline
{\bf 1ifb}
&
583.57
&
131
&
-40.5
&
79.39 (70)
&
Ca binding protein
\\ \hline
{\bf 3cln}
&
835.80
&
143
&
-309.9
&
72.03 (70)
&
ribonuclease
\\ \hline
{\bf 3rn3}
&
1018.66
&
124
&
-88.8
&
72.58 (69)
&
electron transport
\\
\hline
\end{tabular}

\end{center}
\caption{This table gives some statistics of our computed fittest
$H/P$ protein sequences for the 34 empirical protein 3D structures
chosen from PDB.  Column 1 contains the PDB name of the native protein
sequence, where the proteins from Kleinberg~\cite{Kleinberg:1999:EAP} are
in bold.  Column 2 shows the length normalized solvent accessibility,
i.e., $\sum_{i \in H(x)} s_i/n$, of the computed fittest protein
sequence. Column 4 lists the protein sequence length $n$, i.e., the
count of amino acids.  Column 4 provides the computed optimal ratio of
the $\alpha$ and $\beta$ parameters (see text).  Column 5 displays the
percentage of similarity of the computed fittest protein sequence to
the native protein sequence.  Column 6 describes the function of the
native protein.}
\label{table-results}
\end{table}

\begin{figure}[thb]
\centerline{\psfig{figure=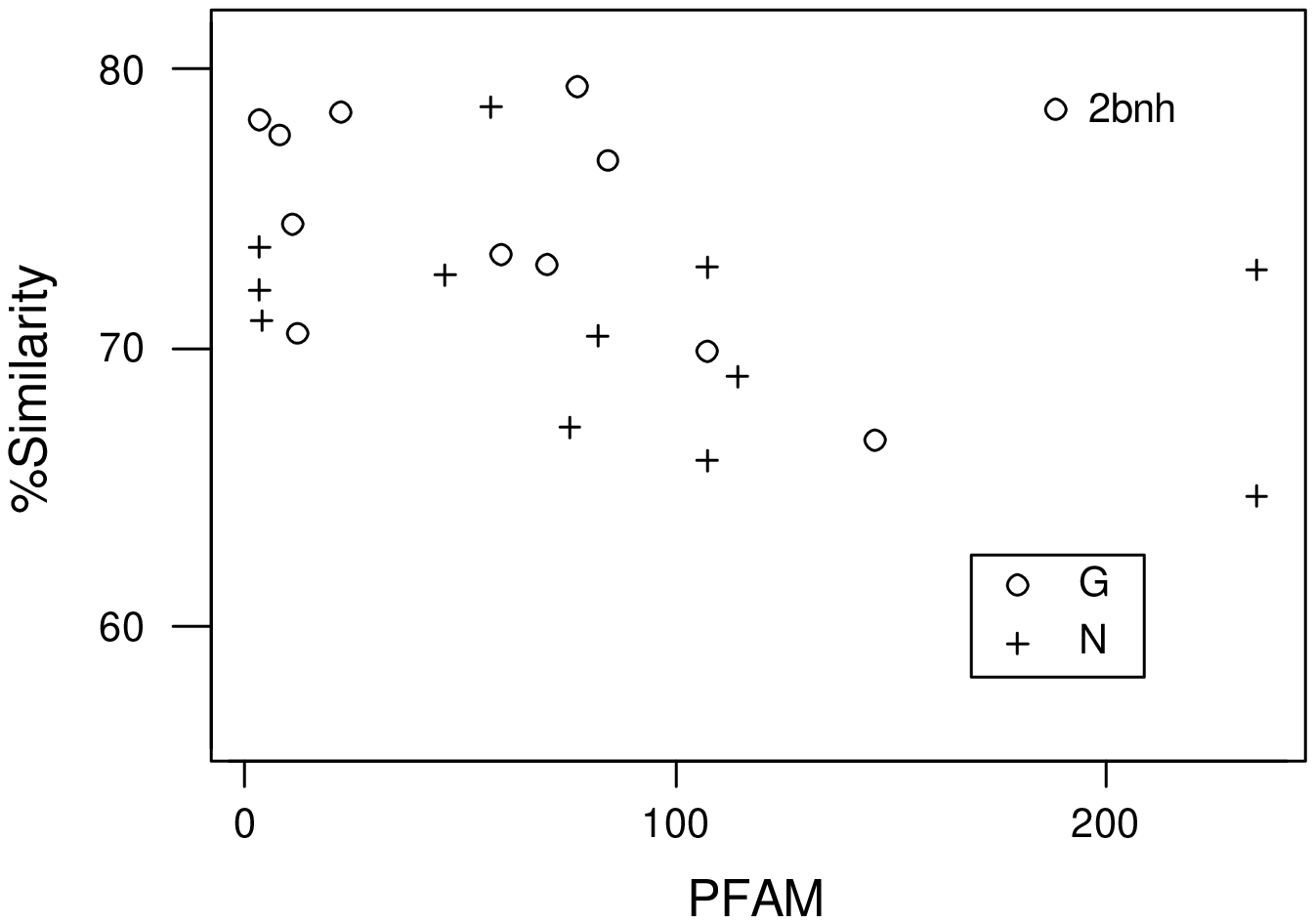}}
\caption{This plot displays the relationship between the percent
similarity of computed fittest protein sequences to native protein
sequences versus the PFAM family size of native protein
sequences. Globular native proteins are shown as open circles while
nonglobular ones are shown as crosses.}
\label{fig:similarity_PFAM}
\end{figure}

To demonstrate our algorithms, we chose 34 proteins with known 3D
structures from the Protein Data Bank (PDB)
at~\texttt{http://www.rcsb.org/pdb}. These 3D structures included 8
from Kleinberg's study~\cite{Kleinberg:1999:EAP} but excluded the protein
fragments and multimeric proteins used in that study. The chosen 3D
structures were then represented by centroids for each side chain
calculated from the coordinates of each atom in the side chain; in the
case of 3D structures solved by NMR, hydrogen atoms were included into
centroid calculations.  For glycine, the centroid was taken to be the
position of $C_\alpha$. For each side chain, the area of solvent
accessible surface was computed via the Web interface of the ASC
program with default parameters \cite{Eisenhaber:1993:ISA}.  In accordance
to the GC model, each of the chosen native protein sequences was
converted into a binary $H/P$ sequence following
Sun~\etal~\cite{Sun:1995:DAA}, where A, C, F, I, L, M, V, W, Y are $H$, and
the other amino acids are $P$.

We used Equation~\ref{eq-phi-definition} in the GC model to calculate
fitness values of protein sequences to determine minimal energy values
and consequently to compute the ``fittest'' protein sequences. These
fitness values consist of two terms in
Equation~\ref{eq-phi-definition}.  The first term accounts for the
idea that hydrophobic residues tend to cluster together due to
stacking forces from the solvent.  The second term accounts for the
idea that hydrophobic residues tend to avoid solvent accessible
surfaces of the molecule. The arbitrary parameters $\alpha$ and
$\beta$ represent scaling factors for the relative importance of these
two tendencies. We expected the appropriate ratio of these two values
to depend on the type of a protein (globular, nonglobular, and
membrane) and the length of the protein. Therefore, we optimized the
scale of the two parameters to find a ratio that maximizes the
similarity of a fittest protein sequence to the native protein
sequence, similarly in spirit to Kleinberg's scaling
algorithm~\cite{Kleinberg:1999:EAP}.

Results of this optimization are shown in Table~\ref{table-results}.
As anticipated, our algorithms computed fittest protein sequences that
are closer to native protein sequences than found by
Kleinberg~\cite{Kleinberg:1999:EAP}, whose results are shown in parenthesis
in Table~\ref{table-results}.  (Note that protein 1aaj is an exception
to this improvement on proximity---perhaps due to the fact that the
input data are not exactly the same.)  However, proximity to native
protein sequences is not a good proxy for biological relevance of the
algorithms. Determination of protein 3D structure involves an energy
landscape given by a statistical thermodynamic energy function
$E(\SSS,\DD)$ where $\SSS$ is the set of amino acid sequences and
$\DD$ is the set of possible folded 3D structures. On the one hand,
for a fixed protein sequence $x$, the 3D structure $D$ is determined
by a temperature-dependent folding process that minimizes $E(x,D)$
over $\DD$.  {\em Ab initio} solutions to this problem for protein
sequences of practical length currently do not exist.  On the other
hand, for a fixed 3D structure $D$, no known thermodynamic reasons
connect the folding process with a protein sequence $x$ that minimizes
$E(x,D)$ over $\SSS$.  However, we can invoke evolutionary processes
as possibly selecting for those protein sequences that produce the
most stable 3D structures, i.e., those with the lowest $E(x,D)$, at a
given temperature. In other words, given a suite of protein sequences
that fold into a particular 3D structure to perform a biological
function, there might be selection for the most thermodynamically
stable protein sequences.

{From} the view point of evolutionary selection, the difference
between computed fittest protein sequences and native protein
sequences may be attributed to three factors. First, the GC toy
thermodynamic model is inappropriate. Second, the biological function
of a protein actually requires structural lability. Last, a native
protein is part of a diverse family and other members of the family
lie closer to the computed fittest protein sequence. This last factor
can be augmented by the argument that if a computed fittest protein
sequence exists in nature but is very different from the native
protein sequence, it is likely that many other protein sequences (thus
a diverse family of protein sequences) exist in nature and fold into
the same or similar 3D structures. All of these factors are likely to
play in the data shown in Table~\ref{table-results}.  However, we
conjectured a significant relationship between a computed fittest
protein sequence's similarity to a native protein sequence and the
diversity of the native protein in nature.  Such a relationship would
be highly intriguing biologically. We examined this conjecture by
assessing the diversity of native proteins using the database PFAM
at~\texttt{http://pfam.wustl.edu}, which is a database of protein
families determined through Hidden Markov Models
\cite{Bateman:2000:PDP}. The database contained information on the putative
family size of 25 of our 34 chosen native
proteins. Figure~\ref{fig:similarity_PFAM} shows the plot of the
percent similarity of computed fittest protein sequences to native
protein sequences versus the PFAM family size of native proteins. In
the figure, globular proteins are shown as circles and nonglobular
ones as crosses. There is a negative linear trend as suggested by our
conjecture. Linear regression is nearly significant at 0.05 level with
$p = 0.088$. The figure shows three outliers, 2bnh, 2stv, and 8cho. Of
these, 2bnh is an exceedingly strange 3D structure with a protein
sequence of alpha helixes forming a horseshoe shaped sheet, resulting
in a 3D structure that is very deviant from globular proteins which
are the genesis of the original thermodynamic model. Leaving out this
outlier results in a significant linear regression with $p = 0.015$.

There is still considerable uncertainty about the appropriateness of
the GC toy model. The average percentage of the hydrophobic residues
is 42\% in the native protein sequences compared to 35\% in the
computed fittest protein sequences. More importantly, the standard
deviation of the percentage of hydrophobic residues is 0.054 in the
native protein sequences compared to 0.143 for the computed fittest
protein sequences. Thus, the percentage of hydrophobic residues is
relatively constant in the native protein sequences, reflecting
perhaps a functional need or unknown structural factors. In contrast,
the percentage of hydrophobic residues in a computed fittest protein
sequence tends to vary depending on the 3D structure. This suggests
that it might be important to introduce a hydrophobic residue
percentage constraint into optimization algorithms in the future as
suggested in the sliding algorithm of
Kleinberg~\cite{Kleinberg:1999:EAP}. Nevertheless, our preliminary results
show that even such a simplified toy model might be useful for
exploratory investigations of protein evolution especially when
coupled to computationally efficient algorithms to allow systematic
investigation of the roughly 13,000 empirical protein 3D
structures. We are currently planning a large-scale analysis of
further empirical protein 3D structures; the results will be reported
in a subsequent paper.

\ifabs{}{
\acknowledgments
\dobib
}


\begin{thebibliography}{10}

\bibitem{Atkins:1999:IPF}
J.~Atkins and W.~E. Hart.
\newblock On the intractability of protein folding with a finite alphabet of
  amino acids.
\newblock {\em Algorithmica}, 25(2-3):279--294, 1999.

\bibitem{Babajide:1997:NNP}
A.~Babajide, I.~Hofacker, M.~Sippl, and P.~Stadler.
\newblock Neutral networks in protein space: {A} computational study based on
  knowledge-based potentials of mean force.
\newblock {\em Folding and Design}, 2:261--269, 1997.

\bibitem{Banavar:1998:SBD}
J.~Banavar, M.~Cieplak, A.~Maritan, G.~Nadig, F.~Seno, and S.~Vishveshwara.
\newblock Structure-based design of model proteins.
\newblock {\em Proteins: Structure, Function, and Genetics}, 31:10--20, 1998.

\bibitem{Bateman:2000:PDP}
A.~Bateman, E.~Birney, R.~Durbin, S.~R. Eddy, K.~L. Howe, and E.~L.~L.
  Sonnhammer.
\newblock {PFAM}-- {A} database of protein domain family alignments and {HMMs}.
\newblock {\em Nucleic Acids Research}, 28:263--266, 2000.

\bibitem{Berger:1998:PFH}
B.~Berger and T.~Leighton.
\newblock Protein folding in the hydrophobic-hydrophilic {(HP)} model is
  {NP}-complete.
\newblock {\em Journal of Computational Biology}, 5(1):27--40, 1998.

\bibitem{Crescenzi:1998:CPF}
P.~Crescenzi, D.~Goldman, C.~Papadimitriou, A.~Piccolboni, and M.~Yannakakis.
\newblock On the complexity of protein folding.
\newblock {\em Journal of Computational Biology}, pages 423--466, 1998.

\bibitem{Deutsch:1996:NAP}
J.~M. Deutsch and T.~Kurosky.
\newblock New algorithm for protein design.
\newblock {\em Physical Review Letters}, 76:323--326, 1996.

\bibitem{Dill:1995:PPF}
K.~A. Dill, S.~Bromberg, K.~Yue, K.~Fiebig, D.~Yee, P.~Thomas, and H.~S. Chan.
\newblock Principles of protein folding --- {A} perspective from simple exact
  models.
\newblock {\em Protein Science}, 4:561--602, 1995.

\bibitem{Drexler:1981:MEA}
K.~E. Drexler.
\newblock Molecular engineering: An approach to the development of general
  capabilities for molecular manipulation.
\newblock {\em Proceedings of the National Academy of Sciences of the U.S.A.},
  78:5275--5278, 1981.

\bibitem{Eisenhaber:1993:ISA}
F.~Eisenhaber and P.~Argos.
\newblock Improved strategy in analytic surface calculation for molecular
  systems: Handling of singularities and computational efficiency.
\newblock {\em Journal of Computational Chemistry}, 14(N11):1272--1280, 1993.

\bibitem{Eisenhaber:1995:DCL}
F.~Eisenhaber, P.~Lijnzaad, P.~Argos, C.~Sander, and M.~Scharf.
\newblock The double cube lattice method: Efficient approaches to numerical
  integration of surface area and volume and to dot surface contouring of
  molecular assemblies.
\newblock {\em Journal of Computational Chemistry}, 16(N3):273--284, 1995.

\bibitem{Gabow:1991:APR}
H.~N. Gabow.
\newblock Applications of a poset representation to edge connectivity and graph
  rigidity.
\newblock In {\em Proceedings of the 32nd Annual IEEE Symposium on Foundations
  of Computer Science}, pages 812--821, 1991.

\bibitem{Garey:1979:CIG}
M.~Garey and D.~Johnson.
\newblock {\em Computers and Intractability: {A} Guide to the Theory of
  {NP}-Completeness}.
\newblock Freeman, New York, NY, 1979.

\bibitem{Goldberg:1988:NAM}
A.~V. Goldberg and R.~E. Tarjan.
\newblock A new approach to the maximum-flow problem.
\newblock {\em Journal of the ACM}, 35(4):921--940, Oct. 1988.

\bibitem{Grotschel:1988:GAC}
M.~Gr{\"{o}}tschel, L.~Lov\'{a}sz, and A.~Schrijver.
\newblock {\em {Geometric Algorithms and Combinatorial Optimization}}, volume~2
  of {\em Algorithms and Combinatorics}.
\newblock Springer-Verlag, New York, NY, 1988.

\bibitem{Hart:1997:CCS}
W.~E. Hart.
\newblock On the computational complexity of sequence design problems.
\newblock In {\em Proceedings of the 1st Annual International Conference on
  Computational Molecular Biology}, pages 128--136, 1997.

\bibitem{Kimura:1983:NTM}
M.~Kimura.
\newblock {\em The Neutral Theory of Molecular Evolution}.
\newblock Cambridge University Press, Cambridge, United Kingdom, 1983.

\bibitem{Kleinberg:1999:EAP}
J.~M. Kleinberg.
\newblock Efficient algorithms for protein sequence design and the analysis of
  certain evolutionary fitness landscapes.
\newblock In {\em Proceedings of the 3rd Annual International Conference on
  Computational Molecular Biology}, pages 226--237, 1999.

\bibitem{Lau:1989:LSM}
K.~F. Lau and K.~A. Dill.
\newblock A lattice statistical mechanics model of the conformational and
  sequence spaces of proteins.
\newblock {\em Macromolecules}, 22:3986--3997, 1989.

\bibitem{Lau:1990:TPM}
K.~F. Lau and K.~A. Dill.
\newblock Theory for protein mutability and biogenesis.
\newblock {\em Proceedings of the National Academy of Sciences of the U.S.A.},
  87:638--642, 1990.

\bibitem{Lawler:1972:PCB}
E.~L. Lawler.
\newblock A procedure for computing the {$k$} best solutions to discrete
  optimization problems and its application to the shortest path problem.
\newblock {\em Management Science}, 18:401--405, 1972.

\bibitem{Lipman:1991:MNS}
D.~Lipman and W.~Wilbur.
\newblock Modeling neutral and selective evolution of protein folding.
\newblock {\em Proceedings of Royal Society of London Series B}, 245:7--11,
  1991.

\bibitem{Merz:1994:PFP}
K.~M. Merz and S.~M.~L. Grand, editors.
\newblock {\em The Protein Folding Problem and Tertiary Structure Prediction}.
\newblock Birkhauser, Boston, MA, 1994.

\bibitem{Micheletti:1998:DPH}
C.~Micheletti, F.~Seno, A.~Maritan, and J.~Banavar.
\newblock Design of proteins with hydrophobic and polar amino acids.
\newblock {\em Proteins: Structure, Function, and Genetics}, 32:80--87, 1998.

\bibitem{Papadimitriou:1982:COA}
C.~H. Papadimitriou and K.~Steiglitz.
\newblock {\em Combinatorial Optimization: Algorithms and Complexity}.
\newblock Prentice-Hall, Upper Saddle River, NJ, 1982.

\bibitem{Picard:1980:SAM}
J.-C. Picard and M.~Queyranne.
\newblock On the structure of all minimum cuts in a network and applications.
\newblock {\em Mathematical Programming Study}, (13):8--16, 1980.

\bibitem{Ponder:1987:TTP}
J.~Ponder and F.~M. Richards.
\newblock Tertiary templates for proteins.
\newblock {\em Journal of Molecular Biology}, 193:63--89, 1987.

\bibitem{Provan:1983:CCC}
J.~S. Provan and M.~O. Ball.
\newblock The complexity of counting cuts and of computing the probability that
  a graph is connected.
\newblock {\em {SIAM} Journal on Computing}, 12(4):777--788, Nov. 1983.

\bibitem{Reidys:1997:GPC}
C.~Reidys, P.~Stadler, and P.~Schuster.
\newblock Generic properties of combinatory maps: Neutral networks of {RNA}
  secondary structures.
\newblock {\em Bulletin of Mathematical Biology}, 59:339--397, 1997.

\bibitem{Shakhnovich:1993:NAD}
E.~I. Shakhnovich and A.~M. Gutin.
\newblock A new approach to the design of stable proteins.
\newblock {\em Protein Engineering}, 6:793--800, 1993.

\bibitem{Smith:1970:NSC}
J.~M. Smith.
\newblock Natural selection and the concept of a protein space.
\newblock {\em Nature}, 225:563--564, 1970.

\bibitem{Steiner:1986:AGI}
G.~Steiner.
\newblock An algorithm to generate the ideals of a partial order.
\newblock {\em Operations Research Letters}, 5:317--320, 1986.

\bibitem{Sun:1995:DAA}
S.~J. Sun, R.~Brem, H.~S. Chan, and K.~A. Dill.
\newblock Designing amino acid sequences to fold with good hydrophobic cores.
\newblock {\em Protein Engineering}, 8(12):1205--1213, Dec. 1995.

\bibitem{Vazirani:1992:SCT}
V.~V. Vazirani and M.~Yannakakis.
\newblock Suboptimal cuts: Their enumeration, weight and number (extended
  abstract).
\newblock In W.~Kuich, editor, {\em Lecture Notes in Computer Science 623:
  Proceedings of the 19th International Colloquium on Automata, Languages, and
  Programming}, pages 366--377. Springer-Verlag, New York, NY, 1992.

\bibitem{Yue:1992:IPF}
K.~Yue and K.~A. Dill.
\newblock Inverse protein folding problem: Designing polymer sequences.
\newblock {\em Proceedings of the National Academy of Sciences of the U.S.A.},
  89:4163--4167, 1992.

\end{thebibliography}
\end{document}